\documentclass[journal,11pt,onecolumn,draftcls]{IEEEtran}

\usepackage[table]{xcolor}
\usepackage{times}
\usepackage{epsfig}
\usepackage{amsmath}
\usepackage{amsfonts}
\usepackage{graphicx}
\usepackage{amssymb}
\usepackage{amstext}
\usepackage{latexsym}
\usepackage{color}
\usepackage{ifthen}
\usepackage{multirow}
\usepackage{verbatim}
\usepackage{array,tabularx}
\usepackage{arydshln}
\usepackage[mathscr]{euscript}
\usepackage{tikz}
\usepackage{accents}
\usepackage{cite}
\usepackage{cleveref}

\usepackage{soul,framed}

\usepackage{caption}
\usepackage{subcaption}
\usepackage{enumerate}
\usepackage{xcolor}

\usepackage{mathtools}
\usepackage{url}
\usepackage{xparse}
\usetikzlibrary{decorations.pathreplacing}
  \usepackage{makecell}

\usepackage{amsthm}         
\definecolor{arylideyellow}{rgb}{0.91, 0.84, 0.42}
\colorlet{shadecolor}{arylideyellow}
\definecolor{hlcolor}{rgb}{0.91, 0.84, 0.42}

\usepackage{hhline}
\definecolor{xgray}{rgb}{0.4,0.4,0.4}
\newcolumntype{C}[1]{>{\centering\let\newline\\\arraybackslash\hspace{0pt}}m{#1}}

\usetikzlibrary{arrows,positioning,automata,calc}

\DeclarePairedDelimiter\floor{\lfloor}{\rfloor}

\usepackage{pgfplots}
\usepackage{subcaption,graphicx}
\usepackage{tikz}
\usepackage{xcolor}

\usepackage{tikz}
\usetikzlibrary{shapes.geometric}
\usetikzlibrary{shapes,snakes,patterns}
\usetikzlibrary{arrows,positioning,automata,calc}
\usetikzlibrary{plotmarks}

\newtheorem{theorem}{Theorem}

\newtheorem{claim}{Claim}

\newtheorem{example}{Example}
\newtheorem{definition}{Definition}

\newtheorem{remark}{Remark}

\DeclareMathOperator*{\tr}{tr}

\usetikzlibrary{arrows,positioning,automata,calc}
\usetikzlibrary{matrix, positioning,tikzmark,shapes}
\usetikzlibrary{calc, fit}
\usetikzlibrary{shapes.geometric}
\usetikzlibrary{snakes,patterns}
\usetikzlibrary{arrows,automata,calc}
\usetikzlibrary{plotmarks}

\newcommand*{\rowstyle}[1]{
  \gdef\@rowstyle{#1}%
  \leavevmode\@rowstyle
  \ignorespaces
}

\newcolumntype{=}{
  >{\gdef\@rowstyle{}\ignorespaces}%
}
\newcolumntype{+}{
  >{\leavevmode\@rowstyle\ignorespaces}%
}

\newcommand{\spc}{
	\hspace{0.75em}
 }

\newcommand{\myhline}{
\hhline{*{2}{~}*{15}{|-}}
}

\newcommand{\myhlinea}{
\hhline{*{2}{~}*{5}{|-}}
}

\definecolor{ao}{rgb}{0.0, 0.5, 0.0}
\definecolor{amber}{rgb}{1.0, 0.75, 0.0}
\definecolor{capri}{rgb}{0.0, 0.75, 1.0}
\definecolor{chocolate}{rgb}{0.91, 0.41, 0.17}

		\tikzset{
			pobl/.style={
				inner sep=0pt, outer sep=0pt, fill=#1,
			},
			pobl gron/.style n args={2}{
				pobl=#1, rounded corners=#2,
			},
			pics/person/.style n args={2}{
				code={
					\node (-corff) [pobl=#1, minimum width=.25*#2, minimum height=.375*#2,  pic actions,rounded corners=0pt] {};
					\node (-pen) [minimum width=.27*#2, circle, pobl=#1, outer sep=.01*#2, anchor=south,  pic actions] at (-corff.north) {};
					\node (-coes dde) [pobl gron={#1}{0.02*#2}, anchor=north west, minimum width=.12125*#2, minimum height=.25*#2,  pic actions] at (-corff.south west) {};
					\node [pobl=#1, anchor=north, minimum width=.12125*#2, minimum height=.15*#2,  pic actions] at (-coes dde.north) {};
					\node (-coes chwith) [pobl gron={#1}{0.02*#2}, anchor=north east, minimum width=.12125*#2, minimum height=.25*#2,  pic actions] at (-corff.south east) {};
					\node [pobl=#1, anchor=north, minimum width=.12125*#2, minimum height=.15*#2,  pic actions] at (-coes chwith.north) {};
					\node (-braich dde) [pobl gron={#1}{0.02*#2}, minimum width=.075*#2, minimum height=.325*#2, outer sep=.0064*#2, anchor=north west,  pic actions] at (-corff.north east)  {};
					\node [pobl=#1, minimum width=.05*#2, minimum height=.2*#2, outer sep=.0064*#2, anchor=north west,  pic actions] at (-corff.north east) {};
					\node (-braich chwith) [pobl gron={#1}{0.02*#2}, minimum width=.075*#2, minimum height=.325*#2, outer sep=.0064*#2, anchor=north east,  pic actions] at (-corff.north west) {};
					\node [pobl=#1, minimum width=.0375*#2, minimum height=.2*#2, outer sep=.0064*#2, anchor=north east,  pic actions] at (-corff.north west) {};
					\node (-fit person) [fit={(-pen.north) (-braich dde.east) (-coes chwith.south) (-braich chwith.west)}] {};
				},
			},
		}
\tikzset{cross/.style={cross out, draw=red, very thick, minimum size=9*(#1-\pgflinewidth), inner sep=0pt, outer sep=0pt},
cross/.default={2.5pt}}

\usetikzlibrary{fit}
\tikzset{
  comp/.style = {
    minimum width  = 1cm,
    minimum height = 0.5cm,
    text width     = 1cm,
    inner sep      = 0pt,
    text           = lightgray,
    align          = center,
    font           = \tiny,
    transform shape,
    thick
  },
  monitor/.style = {draw = none, xscale = 17/16, yscale = 10/9},
  display/.style = {shading = axis, left color = black!60, right color = black},
  ut/.style      = {fill = gray}
}
\tikzset{
  computer/.pic = {
    \node(-m) [comp, pic actions, monitor]
      {\phantom{\parbox{\linewidth}{\tikzpictext}}};
    \node[comp, pic actions, display] {\tikzpictext};
    \begin{scope}[x = (-m.east), y = (-m.north)]
      \path[pic actions, draw = none]
        ([yshift=2\pgflinewidth]-0.1,-1) -- (-0.1,-1.3) -- (-1,-1.3) --
        (-1,-2.4) -- (1,-2.4) -- (1,-1.3) -- (0.1,-1.3) --
        ([yshift=2\pgflinewidth]0.1,-1);
      \path[ut]
        (-1,-2.4) rectangle (1,-1.3)
        (-0.9,-1.4) -- (-0.7,-2.3) -- (0.7,-2.3) -- (0.9,-1.4) -- cycle;
      \path[pic actions, fill = none]
        (-1,1) -- (-1,-1) -- (-0.1,-1) -- (-0.1,-1.3) -- (-1,-1.3) --
        (-1,-2.4) coordinate(sw)coordinate[pos=0.5] (-b west) --
        (1,-2.4) -- (1,-1.3) coordinate[pos=0.5] (-b east) --
        (0.1,-1.3) -- (0.1,-1) -- (1,-1) -- (1,1) -- cycle;
      \node(-c) [fit = (sw)(-m.north east), inner sep = 0pt] {};
    \end{scope}
  }
}

\usepackage{nomencl}
\makenomenclature

\IEEEoverridecommandlockouts

\begin{document}
\title{Private Information Retrieval from MDS  Coded Data in Distributed Storage Systems}
\author{
\IEEEauthorblockN{Razane Tajeddine\IEEEauthorrefmark{1},
	Oliver W.~Gnilke\IEEEauthorrefmark{1}, and
	Salim El Rouayheb\IEEEauthorrefmark{2}}\\
		\IEEEauthorblockA{\IEEEauthorrefmark{1}
		Department of Mathematics and Systems Analysis\\
		Aalto University, School of Science, Finland\\
		Email: \{razane.tajeddine, oliver.gnilke\}@aalto.fi}\\
        \IEEEauthorblockA{\IEEEauthorrefmark{2}ECE Department, Rutgers University\\
		Email:  salim.elrouayheb@rutgers.edu} \thanks{This paper is an extension to the work which has  been presented in part at the IEEE International Symposium on Information Theory (ISIT) 2016 \cite{tajeddine2016private}. This work was supported in part by NSF Grant CCF 1817635. Furthermore, we gratefully acknowledge financial support by the Academy of Finland through grants \#276031, \#282938, and \#303819 to C. Hollanti, Aalto University, Finland. Part of  this work was  carried out while the first and third authors were with the ECE department at the Illinois Institute of Technology, Chicago. The first two authors would like to thank the Institute for Communications Engineering at the Technical University of Munich for hosting them while part of this work was carried out.  Copyright (c) 2017 IEEE. Personal use of this material is permitted.  However, permission to use this material for any other purposes must be obtained from the IEEE by sending a request to pubs-permissions@ieee.org.}}

\maketitle

\begin{abstract}
The problem of providing privacy, in the private information retrieval (PIR) sense, to users requesting data from a distributed storage system (DSS), is considered. The DSS is coded by an $(n,k,d)$ Maximum Distance Separable (MDS) code to store the data reliably on unreliable storage nodes. Some of these nodes can be spies which report to a third party, such as an oppressive regime, which data is being requested by the user. An information theoretic PIR scheme ensures that a user can satisfy its request while revealing no information on which data is being requested to the nodes. A user can trivially achieve PIR by downloading all the data in the DSS. However, this is not a feasible solution due to its high communication cost. We construct PIR schemes with low download communication cost. When there is $b=1$ spy node in the DSS, in other words, no collusion between the nodes, we construct PIR schemes with download cost $\frac{1}{1-R}$ per unit of requested data  ($R=k/n$ is the code rate), achieving the information theoretic limit for linear schemes. The proposed schemes are  universal since they depend on the code rate, but not on the generator matrix of the code. Also, if $b\leq n-\delta k$ nodes collude, with $\delta=\floor{\frac{n-b}{k}}$, we construct linear PIR schemes with download cost $\frac{b+\delta k}{\delta}$.\end{abstract}

\section{Introduction}

Consider the following scenario. A group of online peers (storage nodes) want to collaborate together to form a peer-to-peer (p2p) distributed storage system (DSS) to   store  and share files  reliably, while  ensuring information theoretic private information retrieval (PIR). The PIR \cite{PIR1995, chor1998private} property allows a user (possibly one of the peers) to download a file while revealing no information about which file is being downloaded. We are mainly motivated by the following two applications: 1)  A DSS that protects users from surveillance and monitoring, for instance from an oppressive  regime. The people (peers) collectively contribute to storing the data and making it pervasively  available online. But, some peers could be  spies for the regime. They could   turn against their ``neighbors" and  report to the oppressor  the identity of  users  accessing some  information deemed to be anti-regime (blogs, photos, videos, etc.), leading to their persecution; 
2) A DSS that protects the personal information of  users, such as gender, age group, disease, etc.,  which can be inferred from their file access history. This information can potentially be used to target them with unwanted advertisement, or even affect them adversarially in other areas, such as applications to health insurance or bank loans. In this respect, the studied  DSS can provide an infrastructure, at least in theory, over which applications, such as  cloud storage and social networking, can be run with a privacy guarantee for the users.

We suppose the DSS is formed of $n$ peers or nodes. Peers  can be temporarily offline or can leave the system at any time. The data is stored redundantly in the system to guarantee its durability and  availability. 
We assume that the DSS uses an $(n,k,d)$ maximum distance separable (MDS) code that can tolerate $n-k$ simultaneous node failures.  A certain number of nodes in the DSS, say $b$, whose identities are unknown to the users or the system,  are spies that collude and  can report the user  requests to the oppressor, or sell this information  to interested third parties.
The user can always achieve PIR by asking to download all the files in the DSS. However, this solution is not feasible due to its high communication cost, and  more efficient solutions have been studied in the PIR literature \cite{beimel2005general, yekhanin2010private, beimel2001information, beimel2000reducing, beimel2002breaking, shah2014one, chan2014private} assuming the data is {\em replicated} in the system.  The next example illustrates our PIR scheme with efficient communication cost that can be run on MDS coded data.\vspace{-1mm}

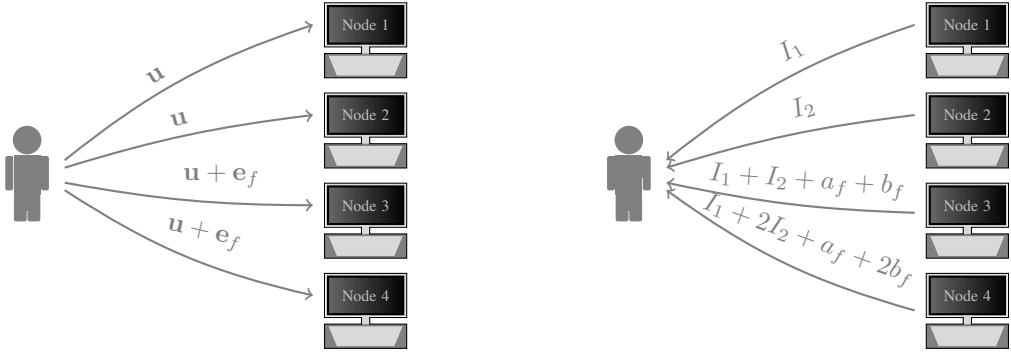
\begin{figure}
\centering
\begin{tikzpicture}
 
 \draw pic at(3,-2) [] {person={gray}{40pt}};
 
  \pic(comp0) at (7.5,0) [
    draw,
    fill = gray!30,
    pic text = {Node 1}
  ]
  {computer};
  
  \pic(comp1) at (7.5,-1.2)[ 
    draw,
    fill = gray!30,
    pic text = {Node 2}
  ]
  {computer};
  
  \pic(comp2) at (7.5,-2.4)[
    draw,
    fill = gray!30,
    pic text = {Node 3}
  ]
  {computer};


  \pic(comp3) at (7.5,-3.6) [
    draw,
    fill = gray!30,
    pic text = {Node 4}
  ]
  {computer};

\draw[->,thick, gray] (3.5,-1.8) to [bend left=10]  node[pos=0.35,above right,gray, rotate=32](r1){\small $\mathbf{u}$}(6.8,0);
\draw[->,thick, gray] (3.5,-1.9) to [bend left=5] node[pos=0.4,above right,gray, rotate=15](r2){\small $\mathbf{u}$}(6.8,-1.2);
\draw[->,thick, gray] (3.5,-2.1) to [bend left=-5] node[pos=0.43,above right,gray](r2){\small $\mathbf{u}+\mathbf{e}_f$}(6.8,-2.4);
 \draw[->,thick, gray] (3.5,-2.2) to [bend left=-10] node[above,xshift=6, yshift=0.01mm,gray, rotate=-13](r3){\small $\mathbf{u}+\mathbf{e}_{f}$} (6.8,-3.6);

 \draw pic at(11,-2) [] {person={gray}{40pt}};
 
  \pic(comp0) at (15.5,0) [
    draw,
    fill = gray!30,
    pic text = {Node 1}
  ]
  {computer};
  
  \pic(comp1) at (15.5,-1.2)[ 
    draw,
    fill = gray!30,
    pic text = {Node 2}
  ]
  {computer};
  
  \pic(comp2) at (15.5,-2.4)[
    draw,
    fill = gray!30,
    pic text = {Node 3}
  ]
  {computer};


  \pic(comp3) at (15.5,-3.6) [
    draw,
    fill = gray!30,
    pic text = {Node 4}
  ]
  {computer};

\draw[->,thick, gray] (14.8,0) to [bend left=-10] node[pos=0.5,above right,gray, rotate=32](r1){\small $I_1$}(11.5,-1.8);
\draw[->,thick, gray] (14.8,-1.2) to [bend left=-5] node[pos=0.5,above right,gray, rotate=19](r2){\small $I_2$} (11.5,-1.9);
\draw[->,thick, gray] (14.8,-2.5) to [bend left=5] node[pos=0,above left,gray, rotate=-4](r2){\small $I_1+I_2+a_f+b_f$} (11.5,-2.1);
\draw[->,thick, gray] (14.8,-3.8) to [bend left=10] node[above,xshift=6, yshift=0.01mm,gray, rotate=-20](r3){\small $I_1+2I_2+a_f+2b_f$} (11.5,-2.2);

\end{tikzpicture}
\caption{\small The user sends queries as specified in Example~\ref{ex:intro} and receives the responses. From the responses, the user can decode $a_f$ and $b_f$, thus decoding the desired file privately.}\vspace{-0.3cm}
\label{fig:ex1}
\end{figure}

\begin{example} \label{ex:intro}  Consider a DSS formed of $n=4$ nodes, as shown in Figure~\ref{fig:ex1}, that stores $m$ files $(a_i,b_i), a_i,b_i\in GF(3^w), i=1,2,\dots, m$. The DSS is coded by an $(n,k,d)=(4,2,3)$ MDS code over $GF(3)$ to store the files. Nodes $1,\dots, 4$ store, respectively, $a_i, b_i, a_i+b_i, a_i+2b_i, i=1,\dots,m$.  Suppose the user is interested in retrieving file $f$, \emph{i.e.}, $(a_f,b_f)$, which can  equally likely be any of the $m$ files. To this end, the user  generates a random vector  $\mathbf{u}=(u_1,\dots,u_m)$ with components chosen independently and uniformly at random from the underlying base field $GF(3)$. It  sends the query vector $\mathbf{q} = \mathbf{u}$ to nodes $1$ and $2$ and $\mathbf{q}=\mathbf{u}+\mathbf{e}_f$ to nodes $3$ and $4$, where $\mathbf{e}_f$ is the all zero vector of length $m$ with a $1$ in the $f^{th}$ entry. Upon receiving the user's request, each node in the DSS returns to the user the projection of all its data on the received query vector. For instance, suppose  that the user wants file $1$. Then, nodes $1, \dots,4$ return the following symbols from $GF(3^w)$, $I_1, I_2, a_f+b_f+I_1+I_2, a_f+2b_f+I_1+2I_2$, where $I_1=\sum_{i=1}^mu_i a_i$ and  $I_2=\sum_{i=1}^mu_i b_i$  are thought of as ``interference" terms. The returned information forms an invertible linear system and the user can decode $a_f$ and $b_f$. Assume that the DSS  contains no colluding nodes, i.e. $b=1$. Then, the proposed  scheme achieves PIR since  the query vector to each node  is statistically independent of the file index $f$. However, if a node, say node $1$,  knows the query vector of another node, say node $3$, it may  be able to pin down which file the user wanted, by computing $\mathbf{e}_f=\mathbf{q}-\mathbf{u}$. However, we assume that a node does not have access to the queries coming to any other nodes, and PIR is indeed achieved here. This PIR scheme downloads $4$ symbols to retrieve a file of size $2$ symbols. We say that the  communication price of privacy $cPoP=4/2=2$ for this scheme, which does not depend on the number of files in the system. 

\end{example}

%

\noindent{\em Replication-based PIR:} PIR was first introduced in the seminal papers of Chor et al.\ in \cite{PIR1995, chor1998private} followed by a significant amount of research in this area \cite{beimel2000reducing, beimel2005general, beimel2002breaking,dvir20142,beimel2001information, yekhanin2008towards,yekhanin2010private}.  
The classical model considers a binary database of length $m$ and a user that wishes  to privately retrieve the value of a  bit (a record)  in it, while minimizing the total communication cost including the upload (query) and download phase. Chor et al.\cite{chor1998private} showed that if there is one server storing the database,  the user has to download the whole database in order to  achieve information theoretic PIR. However, when the database is replicated on $n$ non-colluding (non-cooperating) servers (nodes), they devised a PIR scheme with total, upload and download, communication cost of $O((n^2\log n)m^{1/n})$ and $O(m^{1/3})$ for the special case of $n=2$. 
In the past few years, there has been significant progress in developing  PIR protocols with  total communication cost that is  subpolynomial  in the size of the database  \cite{yekhanin2008towards,efremenko20123, dvir20142}. Moreover, a connection between PIR and blind interference alignment was discussed in \cite{sun2016blind}. PIR in a computational sense was shown to be achievable with a single server (no replication) in \cite{kushilevitz1997replication} assuming the hardness of quadratic residuosity problem. PIR schemes on databases that are replicated but not perfectly synchronized were studied in \cite{7028488}.

\noindent{\em Coded  PIR:} The original model studied in PIR assumes that the entire data  is replicated on each node. PIR on coded data was studied in the literature on Batch Codes \cite{ishai2004batch}, where the data is coded to allow parallel processing leading to  amortizing the PIR communication cost over multiple retrievals. Recently, the PIR problem in  DSSs that use erasure codes was initiated in \cite{shah2014one}, where it was shown that one extra bit of download is sufficient to achieve PIR assuming  the number of servers $n$ to be exponential  in the number of files. Bounds on the information theoretic tradeoff between storage and download communication cost for coded DSSs, for arbitrary number of files $m$, were derived in \cite{chan2014private}. The setting when nodes can be byzantine (malicious) was considered in \cite{augot2014storage} and robust PIR schemes were devised using locally decodable codes. Robust PIR was also studied in \cite{beimel2002robust,tajeddine2017robust}. In \cite{fazeli2015pir}, methods for transforming a  replication-based PIR scheme into a coded-based PIR scheme with the same communication cost, up to a multiplicative constant, were studied. PIR array codes with optimal rate were designed in \cite{blackburn2016pir}.

Following this work in \cite{tajeddine2016private, extended}, the lowest achievable price of privacy for repetition code on $n$ nodes having $m$ files and $b$ colluding nodes was found in \cite{sun2016capacitynoncol, sun2016capacity} to be $\frac{1-(b/n)^m}{1-(b/n)}$ and that of an $(n,k)$-code was found in \cite{banawan2016capacity} to be $\frac{1-(k/n)^m}{1-(k/n)}$. Also, schemes using GRS codes have been constructed in \cite{freij2016private}, and they conjectured that the lowest achievable price of privacy is $\frac{1-(\frac{b+k-1}{n})^m}{1-\frac{b+k-1}{n}}$. That conjecture was then disproved using a counter example in \cite{sun2017private}. Moreover, in \cite{tajeddine2017private}, PIR on coded data such that arbitrary sets of servers collude is studied. In \cite{kumar2016private}, PIR schemes for any arbitrary code were discussed. Some work was also done on symmetric PIR, where the objective is to not only protect the privacy of the user, but also the privacy of the server, such that the user should not get information about files other than the one he wants \cite{wang2017secure}. Also, the capacity of byzantine PIR on replicated storage systems was found in \cite{banawan2017capacity}.


\noindent{\em Contributions:} Motivated by the two DSS applications mentioned earlier, we draw the following distinctions with the previous literature prior to this work on coded PIR \cite{tajeddine2016private}: (i) To the best of our knowledge, all the previous work on coded PIR, except for \cite{chan2014private}, assumes that the code is used to encode together data from different files (records).  However, the model here is different, since in DSS applications only data chunks belonging to the same file are encoded together (as done  in Example~\ref{ex:intro}); (ii) The work in  \cite{chan2014private} studies  fundamental limits on the costs of coded PIR. Here, we provide explicit constructions of PIR schemes with efficient communication cost.

In comparison with the classical literature on  replication-based PIR, we make the following observations: (i)  We focus only on the  number of downloaded symbols in the communication cost of a PIR scheme. This is since the upload query matrices are dependent only on the number of files and not on the size of the file, and typically in DSSs, the size of a file is relatively larger than the total number of files.; (ii) Up to $b$ nodes may collude and share their queries  in the hope of determining the  requested file.

In the model we study, we assume that the MDS code parameters $(n,k,d)$ are given and depend on the desired reliability level for the data. Therefore, they are not design parameters that can be chosen to optimize the efficiency of the PIR scheme. However, the code itself may have to be designed jointly with the PIR scheme.
A PIR scheme incurs many overheads on the DSS, including communication cost, computations \cite{beimel2000reducing}, and connectivity; user contacts  $n$  instead of $k$ nodes,  as seen in Example~\ref{ex:intro}. 
However, we measure here  the efficiency of a PIR scheme only by its total download communication cost, which we refer to as the {\em communication Price of Privacy (cPoP).} A more formal definition of $cPoP$ is given in Definition~\ref{def3} after the model we use has been established. 
The PIR rate is the inverse of the $cPoP$, i.e. the data downloaded from the required file per downloaded symbol.
The following questions naturally arise  here: (1) What is the minimum achievable cPoP for given $n,k$ and $b$? (2) How to efficiently construct codes and PIR schemes that achieve optimal cPoP? (3) Do the code and PIR scheme have to be designed jointly to achieve optimum cPoP? The last question addresses the problem of whether reliability and PIR could be addressed separately in a DSS. Moreover,  it may have practical implications on whether data already existing in coded form needs to be re-encoded  to achieve PIR with minimum cPoP. 

In this paper, we make progress towards answering the last two questions and provide  constructions of efficient PIR schemes for querying MDS coded data.  Specifically, we make the following contributions: (i) For $b=1$, \emph{i.e.}, no colluding nodes, we construct a linear PIR scheme with $cPoP=\frac{1}{1-R}$ ($R=k/n$ is the code rate), thus achieving the lower bound on $cPoP$ for linear schemes in \cite{chan2014private, banawan2016capacity} as $m\to\infty$; (ii) For $2\leq b\leq d-1$, we construct linear PIR schemes with $cPoP= b+k$; (iii) More generally, for $b\leq n-\delta k$, $\delta=\floor{\frac{n-b}{k}}$, we construct linear PIR schemes with $cPoP = \frac{b+\delta k}{\delta}$. While the minimum cPoP in this regime is  unknown, the constructed schemes have a cPoP that does not depend on $m$, the number of files in the system. An important property of  the  scheme for $b=1$ is its {\em universality}. 
It depends only on $n, k,$ and $b$,  but not on the generator matrix of the  code.  Moreover, both of these schemes can be constructed for any given MDS code, \emph{i.e.}, it is not necessary to design the code jointly with the PIR scheme. This implies that $b$ does not have to be a rigid system parameter. Each user can choose their own value of $b$ to reflect  its desired privacy level, at the expense of a higher $cPoP$. The DSS can serve  all the users  simultaneously storing the same encoded data,  \emph{i.e.}, without having to store different encodings for different values of $b$. The construction in \cite{freij2016private} is a generalized version of the earlier scheme presented here. In both schemes, the parity check matrix of the storage system should be known. The two schemes perform equally well, and are in fact identical, for the case of no-collusion ($b=1$) and for the case of $(n-k)$-collusion ($b=n-k$). In the intermediate regime, the generalized scheme in \cite{freij2016private} outperforms our scheme.

  {\small
\begin{table*}[t]
\centering
  \begin{tabular}{|c|c|c|c|c|c|c|}
    \hline 
\rowcolor{gray!25}
   node $1$ & node $2$ & \dots & node $k$ & node $k+1$ & \dots & node $n$ \\\hline
    \tikz[overlay,>=stealth]{\draw [decorate,decoration={brace,mirror}](-12pt,6pt) -- node[left,rotate = 90,yshift=3mm, xshift = 3.5mm,font=\scriptsize]{file $1$} ++(0,-36pt);} 
 $x^1_{11}$ & $x^1_{12}$ & \dots & $x^1_{1k}$ & $\lambda_{1,k+1} x^1_{11} + \dots + \lambda_{k,k+1} x^1_{1k}$ & \dots & $\lambda_{1n} x^1_{11} + \dots + \lambda_{kn} x^1_{1k}$ \\
    \vdots & \vdots & \vdots & \vdots & \vdots & \vdots & \vdots \\
     $x^1_{\alpha1}$ & $x^1_{\alpha2}$ & \dots & $x^1_{\alpha k}$ & $\lambda_{1,k+1} x^1_{\alpha 1} + \dots + \lambda_{k,k+1} x^1_{\alpha k}$ & \dots & $\lambda_{1n} x^1_{\alpha 1} + \dots + \lambda_{kn} x^1_{\alpha k}$ \\\hline
     \tikz[overlay,>=stealth]{\path (-25pt,3pt) -- node {$\vdots$} ++ (0,10pt);} \vdots & \vdots & \vdots & \vdots & \vdots & \vdots & \vdots \\\hline
      \tikz[overlay,>=stealth]{\draw [decorate,decoration={brace,mirror}](-13.5pt,6pt) -- node[left,rotate = 90,yshift=3mm, xshift = 4mm, font=\scriptsize]{file $m$} ++(0,-36pt);}
$x^m_{11}$ & $x^m_{12}$ & \dots & $x^m_{1k}$ & $\lambda_{1,k+1}x^m_{11} + \dots + \lambda_{k,k+1} x^m_{1k}$ & \dots & $\lambda_{1n} x^m_{11} + \dots + \lambda_{kn} x^m_{1k}$ \\
      \vdots & \vdots & \vdots & \vdots & \vdots & \vdots & \vdots \\
     $x^m_{\alpha 1}$ & $x^m_{\alpha 2}$ & \dots & $x^m_{\alpha k}$ & $\lambda_{1,k+1} x^m_{\alpha 1} + \dots + \lambda_{k,k+1} x^m_{\alpha k}$ & \dots & $\lambda_{1n} x^m_{\alpha 1} + \dots + \lambda_{kn} x^m_{\alpha k}$ \\\hline
  \end{tabular}
  \caption{ The layout of the encoded symbols  of the $m$ files in the DSS.}\vspace{-0.5cm}
  \label{general}
\end{table*}
  }


\section{Model}\label{sec:Model}

\noindent{\em Distributed Storage Systems:}
Consider a distributed storage system (DSS) formed of $n$ storage nodes  indexed from $1$ to $n$. The DSS stores   $m$ files, $X^1,\dots,X^m$, of equal sizes. The DSS uses WLOG a systematic\footnote{We focus on systematic codes due to their widespread use in practice. However, our  results still hold for  non-systematic codes.}
$(n,k,d)$ MDS code over $GF(q)$ to store the data redundantly and achieve reliability against  $d-1$ node failures.
We assume that each file, $X^i, i=1,\dots,m$, is divided  into $\alpha$ stripes, and each stripe is divided into $k$ blocks. We represent the file $X^i=[x^i_{lj}], l=1,\dots, \alpha, j=1,\dots, k$, as an $\alpha \times k$ matrix, with symbols from the finite field $GF(q^w)$. We divide the file into stripes to have the number of parts of $X^i$ be divisible by the number of queries and by the number of retrieved symbols per query.

\begin{equation}
{X^i}=\left( {\begin{array}{cccc}
   x^i_{11} & x^i_{12} & \dots & x^i_{1k} \\
   x^i_{21} & x^i_{22} & \dots & x^i_{2k} \\
   \vdots & \vdots & \vdots & \vdots \\
   x^i_{\alpha1} & x^i_{k2} & \dots & x^i_{\alpha k}
\end{array} } \right).
\end{equation}\vspace{0.5em}

Define $\mathscr{X}$ to be the $m\alpha\times k$ matrix denoting all the systematic data in the system, \emph{i.e.}, $$\underset{m\alpha\times k}{\mathscr{X}}=\left(\begin{array}{c}
X^1 \\
X^2 \\
\vdots\\
X^m \\
\end{array}\right).$$

  Each stripe of each file is encoded separately using the same systematic MDS code with a $k\times n$ generator matrix  $\Lambda = [\lambda_{ij}]$ with elements in $GF(q)$. Since the code is systematic, the square submatrix of $\Lambda$ formed of the first $k$ columns is the identity matrix. 
The encoded data, $\mathscr{X}\Lambda$, is stored on the DSS as shown in Table~\ref{general}. We assume that the user knows this layout table, \emph{i.e.}, he/she knows the coding coefficients for each node. We denote by ${\bf{w}}_l\in GF(q^w)^{m\alpha}, l=1,\cdots, n$ the column vector representing all the data on node $l$. 

\noindent{\em PIR:} 
Suppose the user wants file $X^f$, where $f$ is chosen uniformly at random from the set $[m]=\{1,\dots,m\}$. 
To retrieve file $X^f$, the user sends requests to the nodes, among which there are $b$ colluding nodes. The user does not know which nodes are colluding, else, he/she would avoid them.
The goal is to devise a PIR scheme that allows the user to decode $X^f$, while revealing no information, in an information theoretic sense,  about $f$ to the nodes. The colluding nodes can analyze the different requests they receive from the user in order to identify the requested file. However, as explained in the introduction, a node has access to the requests coming to at most $b-1$ other nodes in the system.  Under this setting, we are interested in linear PIR schemes.

\begin{definition}
A  PIR scheme is linear over $GF(q)$, and of dimension $\rho$, if it consists of the following two stages. 

 \noindent{\em 1. Request stage}:   Based on which file the user wants, he/she sends   requests to a subset of nodes in the DSS. The request to node~$l$ takes  the form of a $\rho \times m\alpha$ query matrix $Q_l$ over $GF(q)$.
 
 \noindent{\em 2. Download stage}:  Node $l$ responds by sending the projection of its data onto $Q_l$, \emph{i.e.},   
 \begin{equation}
 R_l = Q_l {\bf{w}}_l\in GF(q^w)^{\rho}. \label{eq:r1}
\end{equation}

\end{definition}

We think  of each query matrix $Q_l$ as formed of $\rho$ sub-queries corresponding to each of its $\rho$ rows. Moreover, we think of  the response of  node $l$ as formed of $\rho$ sub-responses corresponding to projecting the node data on each row of $Q_l$.

\begin{definition}[Information theoretic PIR] A PIR scheme achieves (perfect) information theoretic PIR iff
$H(f|Q_j, j\in\gamma)=H(f)$, for all sets $\gamma \in [n], |\gamma|=b$. 
Here,  $H(\cdot)$ denotes the entropy function.
\end{definition}

The objective is to design a linear PIR scheme that (i) allows the user to decode its requested file $X^f$  and (ii) achieves information theoretic PIR with a low cPoP that does not depend on $m$. In the classical literature on PIR, the communication cost includes  both the number of bits exchanged during the request and download stages. 
However, the query vectors depend only on the number of files in the system, while the response vectors depend on the size of the files, i.e. for a single sub-query, the query vector to a node consists of $m$ symbols in $GF(q)$ while the response vector from one node is $1$ symbol in $GF(q^w)$. In DSSs, and in the information-theoretic reformulation of this problem, the size of the files are assumed to be arbitrarily large, thus making the number of the files negligible with respect to the size of the files \cite{chan2014private}, \emph{i.e.}, $w$ is much larger than $m$. Therefore, the download cost dominates the total communication cost. Hence, we will only consider the download communication cost, which we will refer to as the communication price of privacy (cPoP).

\begin{definition}\label{def3} [cPoP] The communication Price of Privacy (cPoP) of a  PIR scheme  is the ratio of the total number of bits sent from the nodes to the user  during the download stage to size of the requested file. This is the inverse of the PIR rate given in the literature.
\end{definition}
\vspace{4.5em}

\begin{table}
\captionof*{table} {NOMENCLATURE} \label{tab:title} 
\vspace{-1em}
\begin{center}
\begin{tabular}{|c|c|}
\hline
$n$ & Number of nodes in an $(n,k,d)$ MDS code\\\hline
$k$ & Dimension of the codeword in an $(n,k,d)$ MDS code\\\hline
$d$ & Distance of an $(n,k,d)$ code\\\hline
$b$ & Number of colluding nodes \\\hline
$m$ & Number of files \\\hline
$\rho$ & Dimension of the scheme, number of rounds / subqueries / rows in query matrix \\\hline
$r$ & Remainder of the division of $n-k$ by $k$ \\\hline
$\beta$ & Quotient of the division of $n-k$ by $k$ \\\hline
$\alpha$ & Number of subdivisions \\\hline
$\mathbf{u}$ & Random vector of size $m$ \\\hline
$\mathbf{w}_l$ & Data on node $l$\\\hline
$\mathbf{e}_f$ & Indicator vector, the all-zero vector with one $1$ in position $f$ \\\hline
$\mathbf{q}_{l,i}$ & Query vector to Node $l$ in sub-query $i$ \\\hline
$\mathbf{r}_{l,i}$ & Response vector from Node $l$ in sub-query $i$ \\\hline
$Q_l$ & Query Matrix to Node $l$ of dimension $\rho\times m\alpha$ \\\hline
$E_l$ & $0$-$1$ matrix of dimension $\rho\times m\alpha$ \\\hline
\end{tabular}
\end{center}
\end{table}

\section{Main Results}\label{sec:Main}


%
In this section, we state   our two main results. The proof of Theorem~\ref{th:main} is given in Section~\ref{sec:proof}, the proof of Theorem~\ref{th:collusion} is given in Section~\ref{sec:proof2}, and the proof of Theorem~\ref{th:collusion2} is given in Section~\ref{sec:proof3}. 
\begin{theorem} \label{th:main}
Consider a DSS using an $(n,k)$ MDS code over $GF(q)$, with $b=1$, i.e. no collusion between the nodes.  Then, the  linear PIR scheme   over $GF(q)$ described in Section~\ref{sec:constructionb1} achieves perfect PIR with $cPoP=\frac{1}{1-R},$ where $R=k/n$.
\end{theorem}

The existence of PIR schemes over large fields that can achieve  $cPoP= \frac{1}{1-R}$ for $b=1$ follows from Theorem~4 in  \cite{chan2014private}. The scheme in Section~\ref{sec:constructionb1} achieves the optimal $cPoP$ given in \cite{banawan2016capacity} as $m\to \infty$. We prove Theorem~\ref{th:main} by providing an explicit construction of the linear PIR scheme. The proposed PIR construction is over same field over which the code is designed and is universal in the sense that it depends only on the  parameters $n,k$ and $b$ and not on the  generator matrix of the code.

%

\begin{theorem} \label{th:collusion}
Consider a DSS using an $(n,k)$  MDS code over $GF(q)$, with $b$ colluding nodes, $2 \leq b \leq d-1$. Then, there exists an explicit linear PIR scheme over the same field that achieves perfect PIR with $cPoP=b+k$.
\end{theorem}

The next result is a generalization of Theorem~\ref{th:collusion} in which we describe a PIR scheme when $b\leq n-\delta k$, for any $\delta\geq 1$. Theorem~\ref{th:collusion} is a special case of Theorem~\ref{th:collusion2} when $\delta = 1$, but we keep it for a better presentation of the proof. The optimal $cPoP$ for PIR on coded data with colluding nodes is still an open problem.

\begin{theorem} \label{th:collusion2}
For $b\leq n-\delta k$ colluding nodes, with $\delta = \floor{\frac{n-b}{k}}$, we construct an explicit linear PIR scheme with $cPoP = \frac{b + \delta k}{\delta}$.
\end{theorem}



%
%

\begin{figure}
\centering
\begin{tikzpicture}
\pgfplotscreateplotcyclelist{mycolorlist}{%
blue,every mark/.append style={fill=blue!80!black},mark=*\\%
red,every mark/.append style={fill=red!80!black},mark=square\\%
brown!60!black,every mark/.append style={fill=brown!80!black},mark=otimes\\%
black,mark=star\\%
blue,every mark/.append style={fill=blue!80!black},mark=diamond\\%
red,densely dashed,every mark/.append style={solid,fill=red!80!black},mark=*\\%
brown!60!black,densely dashed,every mark/.append style={
solid,fill=brown!80!black},mark=square*\\%
black,densely dashed,every mark/.append style={solid,fill=gray},mark=otimes*\\%
blue,densely dashed,mark=star,every mark/.append style=solid\\%
red,densely dashed,every mark/.append style={solid,fill=red!80!black},mark=diamond*\\%
}
\begin{axis}[
	 xmin=0,
	 xmax=1,
    xlabel={Rate $R=k/n$},
    ylabel={Price of Privacy (cPoP)},
    title={Price of Privacy vs Rate},
    xtick={0, 0.1, 0.2,...,1},
    legend style= {anchor = north west,at={(0,1)}}
  ]

   \addplot+[thick,domain=1/16:15/16,color=chocolate,mark options={chocolate,fill=white},samples=15]{1/(1-x)};  
   
       \addplot +[domain=1/16:15/16,color=blue,mark options={blue,fill=white},samples=15]{1+16*x};
      \addplot +[domain=1/16:15/16,color=green,mark options={green,fill=white},samples=15]{(1+ (floor(15/(16*x)))*(16*x))/(floor(15/(16*x)))};
         \legend{\small Theorem 1 scheme, \small Theorem 2 scheme, \small Theorem 3 scheme.}                         
  \end{axis}
\end{tikzpicture}
\caption{\small cPoP versus rate following the scheme of Theorem~\ref{th:main}, the scheme of Theorem~\ref{th:collusion}, and the scheme for Theorem~\ref{th:collusion2} for $n=16$ and $b=1$}\vspace{-0.3cm}
\label{fig:th1}
\end{figure}
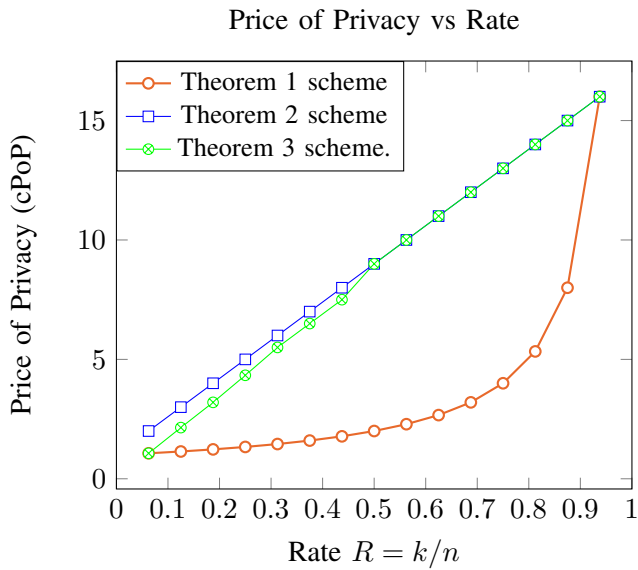

To illustrate the performance stated in the above three theorems, the price of privacy versus the rate of the storage code ($R=\frac{k}{n}$) when using the scheme of Theorem~\ref{th:main}, the scheme of Theorem~\ref{th:collusion}, and the scheme for Theorem~\ref{th:collusion2} for $n=16$ and $b=1$ is shown in Figure~\ref{fig:th1} . We notice that Theorem~\ref{th:main} shows much improvement on Theorem~\ref{th:collusion} and Theorem~\ref{th:collusion2} for $b=1$. We can also see that Theorem~\ref{th:collusion2} improves on Theorem~\ref{th:collusion} when $\delta>1$.

\section {PIR Scheme Construction and proof for $b=1$}\label{sec:Proof}
\subsection{PIR scheme construction for $b=1$}\label{sec:constructionb1}

\ADLinactivate\begin{table}[t]
\def\arraystretch{0.9}
\renewcommand\tabcolsep{3pt}
\centering
\begin{tabular}{p{12pt}r|c|c|c|c|c|c|c|c|c|c|c|c|c|c|c|c|c|}
\multicolumn{1}{c}{~} & \multicolumn{1}{c}{~} & \multicolumn{1}{c}{~} & \multicolumn{1}{c}{~} & \multicolumn{1}{c}{~} & \multicolumn{1}{c}{~} & \multicolumn{1}{c}{~} & \multicolumn{1}{c}{~} & \multicolumn{1}{c}{~} & \multicolumn{1}{c}{~} & \multicolumn{1}{c}{~} & \multicolumn{1}{c}{~} & \multicolumn{1}{c}{~} & \multicolumn{1}{c}{~} & \multicolumn{1}{c}{~} & \multicolumn{1}{c}{~} \\


\multicolumn{2}{c}{~}&\multicolumn{1}{c}{\color{xgray}$1$ \tikz[overlay,>=stealth]{\draw [decorate,decoration=brace](-11pt,9pt) -- node[above,font=\scriptsize]{\color{xgray}Sys. nodes} ++(54.1pt,0);}} & \multicolumn{1}{c}{\color{xgray}$2$} & \multicolumn{1}{c}{\color{xgray} $3$} & \multicolumn{1}{c}{\color{xgray} $4$} & \multicolumn{1}{c}{\color{xgray}  $5$\tikz[overlay,>=stealth]{\draw [decorate,decoration=brace](-9pt,9pt) -- node[above,font=\scriptsize]{\color{xgray} Parity nodes} ++(154pt,0);}} & \multicolumn{1}{c}{\color{xgray}  $6$} & \multicolumn{1}{c}{\color{xgray}  $7$} & \multicolumn{1}{c}{\color{xgray}  $8$} & \multicolumn{1}{c}{\color{xgray}  $9$} & \multicolumn{1}{c}{\color{xgray}  $10$} & \multicolumn{1}{c}{\color{xgray}  $11$} & \multicolumn{1}{c}{\color{xgray}  $12$} & \multicolumn{1}{c}{\color{xgray}  $13$} & \multicolumn{1}{c}{\color{xgray}  $14$} & \multicolumn{1}{c}{\color{xgray}  $15$} \\\myhline

 \parbox[t]{3mm}{\multirow{11}{*}{\rotatebox[origin=c]{90}{\color{xgray} \scriptsize Stripes}}} &
   \color{xgray}  $1$\tikz[overlay,>=stealth]{\draw [decorate,decoration={brace,mirror}](-10pt,6pt) -- node[left,font=\scriptsize]{$r$} ++(0,-25pt);} & \cellcolor{blue!25} $1$& \cellcolor{red!25} $2$ &\cellcolor{green!25} $3$ &\cellcolor{yellow!25} $4$ & &  &  &  &  &  &  &  &  &  & \\\myhline
      
 &    \color{xgray} $2$&\cellcolor{red!25} $2$ & \cellcolor{green!25} $3$ &\cellcolor{yellow!25} $4$ &\cellcolor{blue!25} $1$ & &  &  &  &  &  &  &  &  &  & \\\myhline
       
   &  \color{xgray} $3$&\cellcolor{green!25} $3$ &\cellcolor{yellow!25} $4$ &\cellcolor{blue!25} $1$ &\cellcolor{red!25} $2$ & &  &  &  &  &  &  &  &  &  & \\\myhline
       
     &\color{xgray} $4$\tikz[overlay,>=stealth]{\draw [decorate,decoration={brace,mirror}](-10pt,6pt) -- node[left,font=\scriptsize]{$k$} ++(0,-34pt);}& &  &  &  &\cellcolor{blue!25}  $1$ &\cellcolor{blue!25}  $1$ & \cellcolor{blue!25} $1$ &\cellcolor{blue!25}  $1$ &  &  &  &  &  &  & \\\myhline
        
       &\color{xgray}  $ 5$& &  &  &  &\cellcolor{red!25}  $2$ &\cellcolor{red!25}  $2$ &\cellcolor{red!25}  $2$ &\cellcolor{red!25}  $2$ &  &  &  &  &  &  & \\\myhline
        
       &\color{xgray}  $ 6$& &  &  &  &\cellcolor{green!25}  $3$ &\cellcolor{green!25}  $3$ &\cellcolor{green!25}  $3$ &\cellcolor{green!25}  $3$ &  &  &  &  &  &  & \\\myhline
        
       &\color{xgray}  $ 7$& &  &  &  &\cellcolor{yellow!25}  $4$ &\cellcolor{yellow!25}  $ 4$ &\cellcolor{yellow!25}  $4$ &\cellcolor{yellow!25}  $4$ &  &  &  &  &  &  & \\\myhline
        
       &\color{xgray}  $ 8$\tikz[overlay,>=stealth]{\draw [decorate,decoration={brace,mirror}](-10pt,6pt) -- node[left,font=\scriptsize]{$k$} ++(0,-34pt);}& &  &  &  &  &  &  &  &\cellcolor{blue!25} $1$ &\cellcolor{blue!25}  $1$ &\cellcolor{blue!25}  $1$ &\cellcolor{blue!25}  $1$ & & & \\\myhline
        
       &\color{xgray}  $ 9$& &  &  &  &  &  &  &  &\cellcolor{red!25} $2$ &\cellcolor{red!25}  $2$ &\cellcolor{red!25}  $2$ &\cellcolor{red!25}  $2$ & & & \\\myhline
        
       &\color{xgray}  $ 10$& &  &  &  &  &  &  &  &\cellcolor{green!25} $3$ &\cellcolor{green!25}  $3$ &\cellcolor{green!25}  $3$ &\cellcolor{green!25}  $3$ & & & \\\myhline
        
       &\color{xgray}  $ 11$& &  &  &  &\tikz[overlay,>=stealth,xgray]{\draw [decorate,decoration={brace,mirror}](-7pt,-6pt) -- node[below,font=\scriptsize]{$k$} ++(54pt,0);} &  &  &  &\cellcolor{yellow!25} $4$\tikz[overlay,>=stealth]{\draw [xgray,decorate,decoration={brace,mirror}](-10pt,-6pt) -- node[below,font=\scriptsize]{$k$}  ++(56.7pt,0);} &\cellcolor{yellow!25}  $4$ &\cellcolor{yellow!25}  $4$ & \cellcolor{yellow!25} $4$ & & & \\\cline{3-17}
\end{tabular}
\captionsetup{aboveskip=20pt}
\caption{\small  Example of the retrieval pattern for   $(n,k,d)=(15,4,12)$. The $\alpha\times n$ entries of the table correspond to the $\alpha\times n$ coded symbols of the wanted file. All entries with same number, say $j$ (also given the same color) are privately retrieved in the $j^{th}$ sub-query. Note that there are $k=4$ nodes, including the last $r=3$ nodes,  in every sub-query, that do not have any retrieved symbols. The responses of these nodes are used to decode the ``interference" from all the files, needed to confuse the nodes about what is being requested. This interference is then cancelled out from the other sub-responses in order to decode the desired file symbols in  each sub-query.}\vspace{-.5cm}
\label{Ret}
\end{table}

We describe here the PIR scheme referred to in Theorem~\ref{th:main}. We assume WLOG that the MDS code is systematic.
The PIR scheme uses the number of stripes  $\alpha = d-1$ and the dimension
  $\rho=k$.\footnote{The  parameters can be optimized to  $\alpha=\frac{LCM(k,d-1)}{k}$ and $\rho= \frac{LCM(k,d-1)}{d-1}$, as was done in Example~\ref{ex:intro}. But to simplify notation, we will take $\alpha = d-1$ and $\rho=k$.}
We write $\alpha = \beta k + r$ where, $\beta$ and $r$ are integers and  $0\leq r <k$ and $\beta \geq 0$.

The scheme consists of the user 
sending a $\rho\times m\alpha$ query matrix $Q_l$ to each node $l, l=1,\dots,n$. To form the query matrices, the user generates a $\rho\times m\alpha$ random matrix $U=[u_{ij}]$, whose elements are chosen uniformly at random from $GF(q)$, the same field over which the MDS code is defined. The query matrices have the following structure:
\begin{align}
Q_l&= U+E_{f,l},\, \quad l=1, \dots, n-r,\label{eq:query1}\\
Q_l&=U,\,\quad \quad \quad \ l=n-r+1,\dots, n\label{eq:query2}.
\end{align}
$U$ is  the random component of the query aimed at confusing the nodes about the request, whereas $E_{f,l}$ is a deterministic matrix that depends on the index $f$ of the requested file. The matrices $E_{f,l}$ add parts of the file $X^f$ that is being retrieved to the responses of the nodes. The user can download $n-k$ symbols privately per sub-query, so the matrices $E_{f,l}$ add a symbol to the responses of $n-k$ of the nodes per sub-query. In this scheme, the user retrieves $r$ symbols from the systematic nodes, and $\beta k$ symbols from the parity nodes. Moreover, the retrieved symbols should not be redundant. 
The matrices $E_{f,l}$ are $0$-$1$ matrices of dimensions $\rho\times m\alpha$, every row corresponds to a sub-query and every column corresponds to a stripe of a file. A ``$1$'' in the $(i,j)^{th}$ position of $E_{f,l}$ implies that, during the $i^{th}$ sub-query, the $j^{th}$ symbol on node $l$ is being retrieved privately. The matrices $E_{f,l}$ are designed such that the following conditions hold: 
 
\begin{enumerate}
\item Each row and column of the matrices $E_{f,l}$ contains at most one $1$. The restriction on rows guarantees that we receive one coded symbol from a node, instead of the sum of several symbols. The column condition ensures that every symbol is only retrieved once, and thus, no retrieved symbol is redundant.
\item In each sub-query a $1$ is added to the queries of exactly $n-k$ nodes, \emph{i.e.}, for $n-k$ of the matrices $E_{f,l}$ the $i^{th}$ row contains a $1$. This allows the user to decode a codeword from the MDS storage code, since $k$ symbols are not altered, and subsequently decode $n-k$ symbols of the file $X^f$.
\item If $j$ is the index of a stripe of the requested file $f$ then exactly $k$ of the matrices $E_{f,l}$ contain a $1$ in column $j$. This ensures that we retrieve exactly $k$ MDS coded symbols per row, which are needed to recover the original stripe.
\end{enumerate}
Based on these desired retrieval patterns, we choose 

\ADLactivate\begin{equation}E_{f,1} = \left[ {\begin{array}{@{}c:c:c:c@{}}
   \multirow{2}{*}{$\mathbf{0}_{k\times (f-1)\alpha}$} & I_{r\times r} & \multirow{2}{*}{$\mathbf{0}_{k\times \beta k}$} & \multirow{2}{*}{$\mathbf{0}_{k\times (m-f)\alpha}$}\\
   & \mathbf{0}_{(k-r)\times r} & &
\end{array} } \right],\label{eq:V1}
\end{equation}
and   $E_{f,l}, l=2,\dots,k$, is obtained from matrix $E_{f,l-1}$ by a single downward cyclic shift of its row vectors.

We divide the first $\beta k$ parity nodes into $\beta$ groups of $k$ nodes each. All nodes in group $s$, i.e., nodes $l$ where $sk+1 \leq l \leq sk+k$, receive the same query matrix, such that
\begin{equation}\label{eq:Vi}
E_{f,l}\! =\! \left[\!\begin{array}[h!]{@{\,}c@{\,}:c@{\,}:c@{\,}}
\mathbf{0}_{k\times(f-1)\alpha+r+(s-1)k}&I_{k\times k}&\mathbf{0}_{k\times(\beta-s)k+(m-f)\alpha}\\ 
\end{array}\!\right].
\end{equation}
For the remaining $r$ parity nodes we let
\[E_{f,l}=0 \;, \text{for }\; l>\beta k+k,\]
and they hence all receive the same matrix $U$ as a query.

\begin{claim}\label{claimE}
Conditions 1, 2, and 3 are satisfied in the choice of the $E_{f,l}$ above.
\end{claim}

\begin{proof}
\begin{enumerate}
	\item $E_{f,l}$ has at most one $1$ in each row and column.
	\item For the matrices $E_{f,l}$ sent to the parity nodes, all $\beta k$ of them contain exactly one $1$ in row $i$. 
   
   Since the $k$ matrices $E_{f,l}$ for $1 \leq l \leq k$ sent to the systematic nodes are generated by cyclic row shifts of the matrix in \eqref{eq:V1}, and it contains exactly $r$ rows with a single $1$, we see that $r$ of these matrices contain a $1$ in the $i^{th}$ row. In total we have $\beta k +r = n-k$ matrices $E_{f,l}$ that contain a $1$ in their $i^{th}$ row.

    \item The columns corresponding to the stripes of file $f$ are in the range $(f-1)\alpha < j \leq f\alpha$. For $(f-1)\alpha < j \leq (f-1)\alpha+r$ we see that the $k$ matrices of the form \eqref{eq:V1} contain exactly one $1$ in column $j$. For $(f-1)\alpha+r+(s-1)k < j \leq (f-1)\alpha+r+sk$, $s=1,\cdots,\beta$, the $k$ matrices $E_{f,l}$, for $sk+1 < l \leq sk+k$, contain each one $1$ in column $j$.
\end{enumerate}
\end{proof}

\begin{example}[Retrieval pattern] 

Consider a DSS using an $(n,k,d)=(15,4,12)$ MDS code. Therefore, we have $\rho=k=4$ sub-queries to each node. Also, the number of stripes is $\alpha = d-1 = 11$. This gives $\beta = 2$ and $r=3$. Table~\ref{Ret} gives the retrieval pattern of the PIR scheme, \emph{i.e.}, which file symbols are retrieved in each sub-query.
The 11x15 entries in the table represents all the symbols of the desired file with each node being a column. The numbers (alternatively colors) in each entry indicate in which sub-query the specific symbol is retrieved.
\end{example}

%
\ADLinactivate\begin{table}[t]
\centering
\begin{tabular}{p{2pt}r|c|c|c|c|c|}
\multicolumn{1}{c}{~}&\multicolumn{1}{c}{~}&\multicolumn{1}{c}{~}&\multicolumn{1}{c}{~}&\multicolumn{1}{c}{~}&\multicolumn{1}{c}{~}&\multicolumn{1}{c}{~}\\

\multicolumn{2}{c}{~}&\multicolumn{1}{c}{\color{xgray}\small $1$ \tikz[overlay,>=stealth]{\draw [decorate,decoration=brace](-13.5pt,9pt) -- node[above,font=\scriptsize]{\color{xgray}Sys. nodes} ++(40pt,0);}} & \multicolumn{1}{c}{\color{xgray}\small $2$} & \multicolumn{1}{c}{\color{xgray}\small $3$} & \multicolumn{1}{c}{\color{xgray}\small $4$} & \multicolumn{1}{c}{\color{xgray} \small $5$\tikz[overlay,>=stealth]{\draw [decorate,decoration=brace](-50pt,9pt) -- node[above,font=\scriptsize]{\color{xgray} Parity nodes} ++(55pt,0);}} \\\myhlinea

\parbox[b]{2mm}{\multirow{3}{*}{\rotatebox[origin=c]{90}{\color{xgray} \scriptsize Stripes}}} 
 &
   \color{xgray}  \small  $1$ &\cellcolor{blue!25} \small $1$ & \cellcolor{red!25} \small $2$ & & &\\\myhlinea
      
 &    \color{xgray} \small $ 2$& &  &\cellcolor{blue!25} \small $1$ &\cellcolor{blue!25} \small $1$& \\\myhlinea
       
   &  \color{xgray} \small $ 3$& & &\cellcolor{red!25} \small $2$ &\cellcolor{red!25} \small $2$ & \\\cline{3-7}
\end{tabular}
\captionsetup{aboveskip=10pt}
\caption{\small Retrieval pattern for a (5,2,4) code.}\vspace{-0.3cm}
\label{Ret2}
\end{table}

\begin{example}[Decoding]\label{ex:code} Now consider another example with $(n,k,d) = (5,2,4)$ with generator matrix 
$\Lambda = \left( {\begin{array}{ccccc}
   1 & 0 & 1 & 1 & 1 \\
   0 & 1 & 1 & 2 & 3
\end{array} } \right)$, over $GF(5)$.
Suppose the DSS stores $m=3$ files, $X^1,X^2,X^3$.
%
%
%
%
Our goal is to construct a linear  scheme  that achieves perfect PIR against $b=1$, with $cPoP = \frac{1}{1-R} = \frac{5}{3}$. The construction above gives $\alpha = d-1=3$ and $\rho=k=2$. Thus, a file $X^i$ has the following array structure,
$${X^i} =\left( {\begin{array}{cccc}
   x^i_{11} & x^i_{12} \\
   x^i_{21} & x^i_{22} \\
   x^i_{31} & x^i_{32}
\end{array} } \right)\vspace{0.5em}.$$ Therefore, we get $\beta=1$ and $r=1$.
Suppose  WLOG that the user wants file $X^1$, \emph{i.e.}, $f=1$. The user generates an $2\times 9$ random matrix $U=[u_{ij}]$, whose elements are chosen uniformly at random from $GF(5)$. For the nodes  $1,\dots, 4$, the query matrix $Q_l = U+E_{1,l}$, and $Q_5=U$. Therefore, following \eqref{eq:query1}, \eqref{eq:query2}, \eqref{eq:V1}, \eqref{eq:Vi} and Table~\ref{Ret2} we have

{\small
\begin{align*}
Q_1 &= \left[ {\begin{array}{@{}c@{\spc}c@{\spc}c@{\spc}c@{\spc}c@{\spc}c@{\spc}c@{\spc}c@{\spc}c@{}}
u_{11}+{\color{red}1} & u_{12} & u_{13} & u_{14} & u_{15} & u_{16} & u_{17} & u_{18} & u_{19} \\
u_{21} & u_{22} & u_{23} & u_{24} & u_{25} & u_{26} & u_{27} & u_{28} & u_{29}
\end{array} } \right], \\\\
Q_2 &= \left[ {\begin{array}{@{}c@{\spc}c@{\spc}c@{\spc}c@{\spc}c@{\spc}c@{\spc}c@{\spc}c@{\spc}c@{}}
u_{11} & u_{12} & u_{13} & u_{14} & u_{15} & u_{16} & u_{17} & u_{18} & u_{19} \\
u_{21}+{\color{red}1} & u_{22} & u_{23} & u_{24} & u_{25} & u_{26} & u_{27} & u_{28} & u_{29}
\end{array} } \right], \\\\
Q_3 &= \left[ {\begin{array}{@{}c@{\spc}c@{\spc}c@{\spc}c@{\spc}c@{\spc}c@{\spc}c@{\spc}c@{\spc}c@{}}
u_{11} & u_{12}+{\color{red}1} & u_{13} & u_{14} & u_{15} & u_{16} & u_{17} & u_{18} & u_{19} \\
u_{21} & u_{22} & u_{23}+{\color{red}1} & u_{24} & u_{25} & u_{26} & u_{27} & u_{28} & u_{29}
\end{array} } \right], \\\\
Q_4 &= \left[ {\begin{array}{@{}c@{\spc}c@{\spc}c@{\spc}c@{\spc}c@{\spc}c@{\spc}c@{\spc}c@{\spc}c@{}}
u_{11} & u_{12}+{\color{red}1} & u_{13} & u_{14} & u_{15} & u_{16} & u_{17} & u_{18} & u_{19} \\
u_{21} & u_{22} & u_{23}+{\color{red}1} & u_{24} & u_{25} & u_{26} & u_{27} & u_{28} & u_{29}
\end{array} } \right].
\end{align*}
}

The added $1$s in certain positions of the query matrix are due to the addition of the matrix $E_{1,l}$. This construction achieves perfect privacy, since the only information any node $l$ knows about $f$ is through the query matrix $Q_l$, which is random and independent of $f$.  Next, we want to illustrate how  the user can decode the file symbols.
Each node $l$ sends back the length $2$ vector, ${\bf{r}}_l=(r_{l1},r_{l2}) = Q_l {\bf{w}}_l, l=1,\dots,5$, to the user. Recall that ${\bf{w}}_l$ is the data stored on node $l$. Consider the sub-responses of the $5$ nodes to the first sub-query. They form the following linear system:
\begin{align}
  x^1_{11} + I_1&=r_{11} \\   \label{eq:8}
    I_2&=r_{21}\\  
    x^1_{12}+ x^1_{22}+I_1+I_2&=r_{31}\\  \label{eq:10}
    x^1_{12}+2 x^1_{22}+I_1+2I_2&=r_{41}\\ \label{eq:11}
   I_1+3I_2&=r_{51},
\end{align}
where $I_l={\bf{u}}^T_1{\bf{w}}_l, l=1,2$, and  ${\bf{u}}^T_1$ is the first row  of $U$.  

The user can first decode $I_1$ and $I_2$ from \eqref{eq:8} and \eqref{eq:11}. Then, canceling out the values of $I_1$ and $I_2$ from the remaining equations, the user can solve for $ x^1_{11},  x^1_{12}$ and $ x^1_{22}$. Similarly, the user can obtain $ x^1_{21},  x^1_{13}$ and $ x^1_{23}$ from the sub-responses to the second sub-query. This PIR scheme downloads $2$ symbols from each server. Therefore, it has a $cPoP=\frac{10}{6} = \frac{5}{3}$, which matches the bound in Theorem~\ref{th:main}.
\end{example}

\subsection{Proof of Theorem~\ref{th:main}}\label{sec:proof}

The following remarks from coding theory will be used on several occasions. For more background and proofs we refer to \cite{vanlint}.
\begin{remark}\label{remark1}
A linear $[n,k]$ code $C$ is the set of all vectors $C:=\{xG : x \in \mathbb{F}_q^k\} \subseteq \mathbb{F}_q^n$, where $G$ is a generator matrix of the code. Therefore $C$ is a $k$ dimensional subvectorspace of  $\mathbb{F}_q^n$ and any linear combination of codewords in $C$ is again a codeword in $C$.
\end{remark}
\begin{remark}\label{lem:MDS}
The following statements are equivalent.
\begin{enumerate}
\item A $[n,k]$ code $C$ is MDS
\item For any generator matrix $G_C$ of $C$ any $k$ subset of columns is full rank. 
\item The code $C$ can recover from up to $n-k$ erasures in any coordinates.
\end{enumerate}
\end{remark}

We prove Theorem~\ref{th:main} by showing that the scheme described in Section~\ref{sec:constructionb1} has the following properties.

\noindent{\em Decodability:} 
For any sub-query $i$, we sort the nodes into two groups to prove decodability. By the properties of the $E_{f,l}$ in {Claim~\ref{claimE}} exactly $k$ nodes receive only the vector $\mathbf{u}^T_i$, the $i^{th}$ row of $U$, as a query. And the user is aware of the indices of these nodes. For these nodes $l$, the received symbols are given by
\[ r_{li}= \mathbf{u}^T_i \cdot \mathbf{w}_l.\]
Since every stored stripe is a codeword in $C$, by Remark~\ref{remark1}, any linear combination of stripes will be a codeword too. We notice that $r_{li}$ is indeed the $l^{th}$ component of the codeword $r_i'=\mathbf{u}^T_i \cdot (\mathbf{w}_1,\dots,\mathbf{w}_n)$. Since we have $k$ of its components, we can recover the whole vector $r'_i$ by Remark~\ref{lem:MDS}. 

For the other nodes $\ell$, the $i^{th}$ sub-query is of the form $\mathbf{u}^T_i+\mathbf{e}_g$ where $\mathbf{e}_g$ is a standard basis vector, \emph{i.e.}, a single $1$ has been added to the vector $u_i$ in position $g$. The received symbol $r_{\ell i}=\mathbf{u}^T_i \cdot \mathbf{w}_\ell + \mathbf{e}_g \cdot \mathbf{w}_\ell=\mathbf{u}^T_i \cdot \mathbf{w}_\ell + w_\ell(g)$ therefore is the sum of the  $\ell^{th}$ component of $r'_i$ and the $g^{th}$ symbol of $\mathbf{w}_\ell$.  Since we have recovered $r'_i$ from the $k$ unaltered components, we can retrieve $w_\ell(g)$.

Furthermore, the matrices $E_{f,l}$ are designed such that we retrieve exactly $k$ symbols from every coded stripe of the $f^{th}$ file. Using Remark~\ref{lem:MDS} again allows us to retrieve all stripes of file $X^f$ from these $k$ symbols.

{\em Privacy:} Since $b=1$, the only way a node $l$ can learn  information about $f$ is from its own query matrix $Q_i$.  By, construction  $Q_i$ is statistically independent of $f$ and this scheme achieves perfect privacy.

{\em cPoP:} Every node $l\in[n]$ responds with $\rho=k$ symbols. Therefore, the total number of symbols downloaded by the user is $kn$. Therefore,  $cPoP=\frac{kn}{k(n-k)}=\frac{1}{1-R}$.

\section{PIR Scheme construction and Proof for $b\leq d-1$}

\subsection{PIR scheme construction for $b \leq d-1$}\label{sec:constructionb2}

In this section, we will describe the general PIR scheme that achieves $cPOP=b+k$ by specifying the query matrices to each node. This scheme requires $b\leq d-1$.
To simplify the description of the scheme,  we will assume $b=d-1$.\footnote {If $b<d-1$, only $b+k$ nodes, say the first $b+k$, are queried.}
The scheme has dimension $\rho=k$, \emph{i.e.}, it consists of $\rho=k$ sub-queries. Moreover, the scheme requires no subdivisions, \emph{i.e.}, the number of stripes  $\alpha = 1$. Since there are no subdivisions, we simplify further the notation and write $x_{i1}^j = x_i^j$ to denote the $i^{th}$ systematic symbol of file $X^j$, where $j = 1, \dots, m$.
Denote by $f$ the index of the file that  the user wants, \emph{i.e.}, the user wants to retrieve file $X^f$. WLOG, we assume the MDS code is systematic.

In the $i^{th}$ sub-query, $i=1,\dots,k$, the proposed PIR scheme retrieves systematic symbol  $x_i^f$ of the wanted file $X^f$.
So, by the completion of the scheme, the user will have all the $k$  symbols forming the file.  
In sub-query $i$, the user creates $d-1$ random (column) vectors ${{\mathbf{u}}_{1,i}}, \dots, {{\mathbf{u}}_{d-1,i}}$, of dimension $m$ each, whose elements are chosen uniformly at random from $GF(q)$. 
Recall that the generator matrix for any systematic $(n,k,d)$ MDS code is of the form \begin{equation}\label{eq:gen}
\underset{k\times n}{G} = \left[ {\begin{array}{c|c}
	I_{k\times k} & P_{k\times (d-1)} 
	\end{array}} \right],\end{equation}
where $P$ is a $k\times d-1$ matrix describing the parity nodes. A parity check matrix for this code is then given by 
$$ \underset{(d-1)\times n}{H} = \left[ {\begin{array}{c|c}
	-P^T & I_{(d-1)\times (d-1)}
	\end{array}} \right]. $$


Define $U_i$ to be the $m\times (d-1)$ matrix with its columns being the $b = d-1$ random vectors used in sub-query $i$, \emph{i.e.}, $$\underset{{m \times d-1}}{U_i} = \left[
   \mathbf{u}_{1,i}, 
   \mathbf{u}_{2,i}, 
   \dots, 
   \mathbf{u}_{d-1,i} \right].$$

Now for each sub-query the user generates $m$ random codewords in the dual code by multiplying the random matrix $U_i \in GF(q)^{m \times (d-1)}$ by the parity check matrix $\underset{(d-1)\times n}{H} $ to calculate 
$$U_i \underset{(d-1)\times n}{H} = \left[ \mathbf{q}'_{1,i}, \dots, \mathbf{q}'_{n,i}  \right].$$
Note that each row of $U_iH$ is a codeword in the dual of the MDS code used to store the data. 

For $i=1,\dots,k$, let $\mathbf{q}_{l,i}$ be the $i^{th}$ sub-query vector to node $l$ with $l=1,\dots,n$. These query vectors are chosen as follows:
\begin{equation}
	{{\mathbf{q}}_{l,i}} = 
	\begin{cases}
	\mathbf{q}'_{l,i}+\mathbf{e}_f, \quad\quad & \text{if } l = i,\\
	\mathbf{q}'_{l,i}, \quad\quad & \text{otherwise,}
	\end{cases} 
\end{equation}
where $\mathbf{e}_f$ is the standard basis vector with a single $1$ in position $f$.

Therefore, the  response of node $l$ to the $i^{th}$ sub-query, denoted by ${r}_{l,i}$,  is given by \eqref{eq:r1} and can be written as
\begin{equation}\label{eq:res}
{r}_{l,i} = {{\mathbf{q}}_{l,i}}^T {\mathbf{w}}_l,
\end{equation}
where ${\mathbf{w}}_l$ is the vector representing the data stored on node $l$.

We will give an example.
\begin{example}
Next, we illustrate this scheme through an example. Consider a DSS using the following systematic $(5,3,3)$ MDS code with generator matrix 
$$\Lambda = \left[ {\begin{array}{ccccc}
   1 & 0 & 0 & 1 & 1 \\
   0 & 1 & 0 & 1 & 2 \\
   0 & 0 & 1 & 1 & 3
\end{array} } \right].$$
Suppose the system is storing $m=3$ files, $X^1 =  \left[ {\begin{array}{c}
   a_{1} \\
   a_{2} \\
   a_{3}
\end{array} } \right]$,  $X^2 =  \left[ {\begin{array}{c}
   b_{1} \\
   b_{2} \\
   b_{3}
\end{array} } \right]$ and  $X^3 =  \left[ {\begin{array}{c}
   c_{1} \\
   c_{2} \\
   c_{3}
\end{array} } \right]$. Then, the data is stored on the different nodes in the DSS as described in table~\ref{nodiv}.

\begin{table}[h]
\centering
 \begin{tabular}{|c|c|c|c|c|}
\hline
    node 1 & node 2 & node 3 & node 4 & node 5 \\\hline 
    $a_{1}$ & $a_{2}$ & $a_{3}$ & $a_{1} + a_{2} + a_{3}$ & $a_{1} + 2a_{2} + 3a_{3}$ \\\hline
    $b_{1}$ & $b_{2}$ & $b_{3}$ & $b_{1} + b_{2} + b_{3}$ & $b_{1} + 2b_{2} + 3b_{3}$ \\\hline
    $c_{1}$ & $c_{2}$ & $c_{3}$ & $c_{1} + c_{2} + c_{3}$ & $c_{1} + 2c_{2} + 3c_{3}$ \\\hline
   \end{tabular}
  \caption{(5,3,3) DSS}
  \label{nodiv}
\end{table}

Our goal is to construct a linear scheme that achieves perfect PIR against $b=2$ colluding nodes with $cPoP = k+b = 3+b$. The scheme will consist of $\rho=k=3$ sub-queries.

Suppose  WLOG that the user wants file $X^1$, \emph{i.e.}, $f=1$. We will consider the first sub-query and the remaining sub-queries (\emph{i.e.} sub-queries 2 and 3) follow similarly. 
The user creates $2$ random vectors $\mathbf{u}_{1,1}, \mathbf{u}_{2,1}$ of dimension $m=3$ each. $U_1 = \left[ 
\mathbf{u}_{1,1}, 
\mathbf{u}_{2,1}
\right]$.
The dual code will have a generator matrix $$ \underset{(n-k)\times n}{H} = \left[ {\begin{array}{ccccc}
	-1 & -1 & -1 & 1 & 0 \\
	-1 & -2 & -3 & 0 & 1
	\end{array}} \right]. $$


The sub-query vectors to nodes $1$ to $5$ are the following respectively
\begin{align}
\mathbf{q}_{1,1} &= -\mathbf{u}_{1,1} - \mathbf{u}_{2,1}+\left[ {\begin{array} {c}
1 \\ 0 \\ 0
\end{array} } \right], \\
\mathbf{q}_{2,1} &= -\mathbf{u}_{1,1} - 2\mathbf{u}_{2,1}, \\
\mathbf{q}_{3,1} &= -\mathbf{u}_{1,1} - 3\mathbf{u}_{2,1}, \\
\mathbf{q}_{4,1} &= \mathbf{u}_{1,1}, \\
\mathbf{q}_{5,1} &= \mathbf{u}_{2,1}.
\end{align}

Next, we want to show that the user can decode its requested file correctly. The nodes send back the length $3$ vectors, $\mathbf{r}_l=(r_{l,1},r_{l,2},r_{l,3}), l=1,\dots,5$, to the user. Consider the first symbol in each of the vectors $r_{l,1}$,  which  form the following linear system:

\begin{align}
  a_1 - I_{11} - I_{12} &= r_{1,1}  \label{eq:x7} \\   \label{eq:x8}
    -I_{21} - 2I_{22} &= r_{2,1}\\   \label{eq:x9}
    -I_{31} - 3I_{32} &= r_{3,1}\\  \label{eq:x10}
    I_{11} + I_{21} + I_{31} &= r_{4,1}\\  \label{eq:x11}
     I_{12} + 2I_{22} +3 I_{32} &= r_{5,1} 
\end{align}


where $I_{lj}=\mathbf{u}^T_{j,1}\mathbf{w}_{l}$, for $l = 1, 2, 3$ denoting the node index, and $j=1,2$ denoting the random vector.
In analogy with the interference alignment literature \cite{WD09, SR09}, one can think of $a_{1}$ as the signal to be decoded and $I_{11}, I_{12}, I_{21}, I_{22}, I_{31}, I_{32}$ as the interference. 
And we can notice that if we sum up \cref{eq:x7,eq:x8,eq:x9,eq:x10,eq:x11}, we get $a_1$.
This PIR scheme downloads $3$ packets from each server. Therefore, it has a $cPoP=\frac{5\times 3}{3} = 5$.

\end{example}

As mentioned for $b<d-1$ only $b+k$ nodes are queried as shown in the next example. We will revisit this example in the next section and present a more efficient scheme when explaining Theorem~\ref{th:collusion2}.

\begin{example}\label{ex:th2th3}
Consider the (6,2,5) MDS code in table~\ref{tab:exth2}, where $\Lambda = \left[ \begin{array}{cccccc}
1 & 0 & 1 & 1 & 1 & 1 \\
0 & 1 & 1 & 2 & 3 & 4
\end{array}\right]$. The goal here is to construct a linear scheme that achieves perfect PIR against $b=2$ colluding nodes with $cPoP = k+b = 4$. Assume WLOG the user wants file $X^f$.
The scheme will consist of $\rho=2$ sub-queries. We will consider the first sub-query and the second sub-query follows similarly.

\begin{table}
\centering
  \begin{tabular}{|c|c|c|c|c|c|}
   \hline
    node 1 & node 2 & node 3 & node 4 & node 5 & node 6 \\\hline
    $a_1$ & $b_1$ & $a_1+b_1$ & $a_1 +2b_1$ & $a_1 +3b_1$ & $a_1 +4b_1$ \\\hline 
    $\vdots$ & $\vdots$ & $\vdots$ & $\vdots$ & $\vdots$ & $\vdots$ \\\hline
    $a_m$ & $b_m$ & $a_m + b_m$ & $a_m + 2b_m$ & $a_m+3b_m$ & $a_m+4b_m$ \\\hline
  \end{tabular}
  \caption{(6,2,5) DSS}
  \label{tab:exth2}
\end{table}

In this case, the user will query only $4$ nodes, WLOG the first $4$ nodes, with generator matrix $G = \left[ \begin{array}{cccccc}
1 & 0 & 1 & 1 \\
0 & 1 & 1 & 2 
\end{array}\right]$. As described in section~\ref{sec:constructionb2}, the user creates $2$ random vectors $\mathbf{u}_{1,1}, \mathbf{u}_{2,1}$ of dimension $m$ each, and as in the previous example, forms $U_1 = \left[ 
\mathbf{u}_{1,1}, 
\mathbf{u}_{2,1}
\right]$.
The dual code of $G$ will have a generator matrix $$ \underset{(n-k)\times n}{H} = \left[ {\begin{array}{cccc}
	-1 & -1 & 1 & 0 \\
	-1 & -2 & 0 & 1
	\end{array}} \right]. $$


The sub-query vectors to nodes $1$ to $4$ are the following respectively
\begin{align}
\mathbf{q}_{1,1} &= -\mathbf{u}_{1,1} - \mathbf{u}_{2,1}+\mathbf{e}_f, \\
\mathbf{q}_{2,1} &= -\mathbf{u}_{1,1} - 2\mathbf{u}_{2,1}, \\
\mathbf{q}_{3,1} &= \mathbf{u}_{1,1},  \\
\mathbf{q}_{4,1} &= \mathbf{u}_{2,1}. \\
\end{align}

The nodes will respond to the user by projecting their data on the query matrices. With inspection of the queries, we can see that the user will be able to decode $a_f$ from the first sub-query, and similarly decode $b_f$ from the second sub-query. This achieves a $cPoP=4$.

\end{example}

\subsection{Proof of Theorem~\ref{th:collusion}}\label{sec:proof2}



We prove Theorem~\ref{th:collusion} by showing that the scheme described in Section~\ref{sec:constructionb2} ensures decodability and privacy. The main ingredient in the proof, which makes it different from   the proof of Theorem~\ref{th:main},  is that the scheme does not require the user to decode all the interference terms.
Recall that the user wants to retrieve file $X^f$. We will prove that the user can retrieve $x^f_i$ in the $i^{th}$ sub-query. An alternative proof of Theorem~\ref{th:collusion} is shown in the Appendix (Section~\ref{sec:app}).

\noindent\textit{Decodability:} 

The response of node $l = 1,\dots, n$ to the $i^{th}$ sub-query is given by
\begin{equation}
{r}_{l,i} = 
{\mathbf{q}_{l,i}}^T{\mathbf{w}_l}.
\end{equation}

To decode $x_i^f$, the user sums the responses of all the nodes to the $i^{th}$ sub-query, \emph{i.e.}, it computes $\sum_{l=1}^{n} r_{l,i}$.

\begin{claim}\label{cl} $\sum_{l=1}^{n}r_{l,i} = x_i^f$ \end{claim}

\begin{proof} 
\begin{align}
\sum_{l=1}^{n} r_{l,i} &= \tr{((U_iH)^T\mathscr{X}G)} + e_f^T w_i \label{eq:30}\\
&= \tr{(U_iHG^T\mathscr{X}^T)} + x_i^f \label{eq:31}\\
&=x_i^f. \label{eq:32}
\end{align}

where $\tr(\cdot)$ is the trace operator. Equation~\eqref{eq:30} follows directly from the scheme, equation~\eqref{eq:31} follows from the fact that $\tr(A^TB)=\tr(AB^T)$, and equation~\eqref{eq:32} follows from the fact that $\tr{(U_iHG^T\mathscr{X}^T)} = 0$ since $HG^T = 0$.
\end{proof}


\noindent\textit{Privacy:}
Recall that $f \in [m]$ is the index of the file wanted by the user. Let $S_b$ be a subset of cardinality $b$ of $[n]$ representing the set of $b$ colluding nodes.
We define $Q_{S_b}$ to be the set of  query vectors (or matrices) incoming to the $b$ nodes indexed by $S_b$.
We want to show that when $b$ spies collude, they cannot learn any information about $f$, \emph{i.e.}, $H(f|Q_{S_b}) = H(f)$, for any possible set of colluding nodes $S_b\subset [n], |S_b|=b$.  


\begin{align}
H(f, Q_{S_b}) &= H(f, Q_{S_b}) \\
H(Q_{S_b}) + H(f|Q_{S_b}) &= H(f) + H(Q_{S_b}|f) \\
H(f|Q_{S_b}) &= H(f) + H(Q_{S_b}|f)-H(Q_{S_b})\\
&= H(f) -H(Q_{S_b}) + H(Q_{S_b}|f) \notag  \\
&\text{\quad \quad\quad}-\underbrace{H(Q_{S_b}|f,U_i)}_{=0}\label{eq:p1}\\
& = H(f) -H(Q_{S_b}) + I(Q_{S_b},U_i | f) \\
 &=H(f)-H(Q_{S_b}) + H(U_i|f) \notag \\
 &\text{\quad\quad\quad}-H(U_i|Q_{S_b},f) \label{eq:p2} \\
 &=H(f)-H(Q_{S_b})+H(U_i) \label{eq:p3} \\
 &=H(f). \label{eq:p4}
 \end{align}
 
Where the equality in equation~\eqref{eq:p1} follows from the fact that $H(Q_{S_b}|f,U_i) = 0$, since the query vectors are a deterministic function of  $f$ and $U_i$. Equation~\eqref{eq:p3} follows from $H(U_i|f)=H(U_i)$, since the random matrix  $U_i$ is independent of the file index $f$. Moreover, $H(U_i|Q_{S_b},f)=0$ since by \eqref{eq:q1}, given $f$,  $U_i$ can be decoded from  $Q_{S_b}$ due to the MDS property of the code. Lastly, in \eqref{eq:p4} $H(Q_{S_b})=H(U_i)=m$   follows again from \eqref{eq:q1}  and the MDS property of the code.

\begin{figure}
\centering
\begin{tikzpicture}

\pgfplotscreateplotcyclelist{mycolorlist}{%
blue,every mark/.append style={fill=blue!80!black},mark=*\\%
red,every mark/.append style={fill=red!80!black},mark=square\\%
brown!60!black,every mark/.append style={fill=brown!80!black},mark=otimes\\%
black,mark=star\\%
blue,every mark/.append style={fill=blue!80!black},mark=diamond\\%
red,densely dashed,every mark/.append style={solid,fill=red!80!black},mark=*\\%
brown!60!black,densely dashed,every mark/.append style={
solid,fill=brown!80!black},mark=square*\\%
black,densely dashed,every mark/.append style={solid,fill=gray},mark=otimes*\\%
blue,densely dashed,mark=star,every mark/.append style=solid\\%
red,densely dashed,every mark/.append style={solid,fill=red!80!black},mark=diamond*\\%
}
\begin{axis}[
	 xmin=0,
	 xmax=1,
    xlabel={Rate $R=k/n$},
    ylabel={Price of Privacy (cPoP)},
    xtick={0, 0.1, 0.2,...,1},
    legend style= {anchor = south east,at={(1,0)}}
  ]
  
   \addplot+[domain=1/16:13/16, color=green!50!black, mark = square, mark options={scale = 0.5,green!50!black},samples=13]{3+16*x};               
       \addplot+[domain=1/16:11/16, color=red, mark = triangle, mark options={scale = 1},samples=13]{5+16*x};   
         \addplot +[domain=1/16:13/16,mark options={fill=white},samples=13]{(3+ (floor(13/(16*x)))*(16*x))/(floor(13/(16*x)))};
         \addplot +[domain=1/16:11/16,mark options={fill=white},samples=13]{(5 + (floor(11/(16*x)))*(16*x))/(floor(11/(16*x)))};
         \legend{$b=3$ Theorem~\ref{th:collusion},$b=5$ Theorem~\ref{th:collusion}, $b=3$ Theorem~\ref{th:collusion2},$b=5$ Theorem~\ref{th:collusion2}}                         
  \end{axis}
  \draw[->] (3.2,2.5) -- (2,3.8);
  \node at (1.8,4) {\rotatebox{35}{Increasing $b$}}; 
\end{tikzpicture}
\caption{\small cPoP versus rate when $n=16$ and number of colluding nodes $b = 1, 3, 5$, following the scheme in Theorem~\ref{th:collusion} and Theorem~\ref{th:collusion2}. We notice that as the number of colluding nodes increases, the improvement the scheme in Theorem~\ref{th:collusion2} has over the scheme in Theorem~\ref{th:collusion} grows.}\vspace{-0.3cm}
\label{fig:th2}
\end{figure}
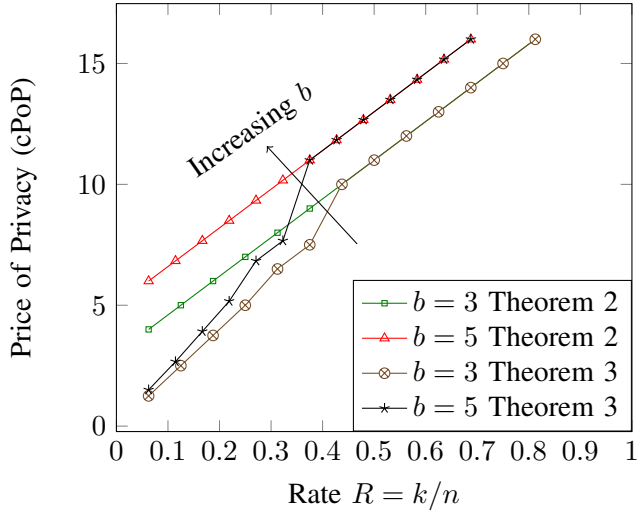

\section{PIR scheme construction for $b\leq n-\delta k$}\label{sec:proof3}

Let $\delta=\floor{\frac{n-b}{k}}$. We can see that for $\delta = 1$, this simplifies to Theorem~\ref{th:collusion}. Figure~\ref{fig:th2} shows a comparison of the construction of Theorem~\ref{th:collusion} and Theorem~\ref{th:collusion2}.

\begin{example}\label{ex:th2th32}
Consider again the (6,2,5) MDS code in table~\ref{tab:exth2} and $b=2$. 
We notice that in example~\ref{ex:th2th3}, we did not use nodes $5$ and $6$, and achieved $cPoP=4$. Now we will show how we can use those nodes and achieve a lower $cPoP=3$. Assume the user wants file $X^f$.

We first choose the last $b=2$ of the parity nodes to be common nodes. Then, we split the rest of the $\delta k = 4$ nodes into $\delta=2$ groups of $k=2$ nodes each. We then consider the two punctured codes, each with a $2\times 4$ generator matrix, that intersect in the common nodes.
Here, we pick the two subcodes consisting of nodes $1, 2, 5, 6$ and $3, 4, 5, 6$, respectively. The punctured codes will have the following generator matrices:

$$G_1 = \left[\begin{array}{c|c}
B_1 & P
\end{array}\right] = \left[ \begin{array}{cc|cc}
1 & 0 & 1 & 1 \\
0 & 1 & 3 & 4
\end{array}\right],$$

and 

$$G_2 = \left[\begin{array}{c|c}
B_2 & P
\end{array}\right] = \left[ \begin{array}{cc|cc}
1 & 1 & 1 & 1 \\
1 & 2 & 3 & 4
\end{array}\right].$$

The user can transform the generator matrices of the punctured codes into systematic form by multiplying by the inverse of the $k\times k = 2\times 2$ matrix formed by the non-common nodes. In this example, we can see that $G_2$ in not in systematic form, so we multiply by the inverse of the $2\times 2$ sub-matrix formed by nodes $3$ and $4$, \emph{i.e.} $\left[ \begin{array}{cc}
1 & 1 \\
1 & 2
\end{array}\right]$ to get
$$G_2 =  \left[ \begin{array}{cccc}
 1 & 0 & -1 & -2 \\
 0 & 1 & 2 & 3
\end{array}\right].$$

The parity check matrices of the $(4,2,3)$ MDS codes generated by $G_1$, and $G_2$ are $$H_1 = \left[\begin{array}{c|c}
-P_1^T & I_{2\times 2}
\end{array}\right] = \left[ \begin{array}{cc|cc}
-1 & -3 & 1 & 0 \\
-1 & -4 & 0 & 1
\end{array}\right],$$ $$H_2 = \left[\begin{array}{c|c}
-P_2^T & I_{2\times 2}
\end{array}\right] =\left[ \begin{array}{cc|cc}
1 & -2 & 1 & 0 \\
2 & -3 & 0 & 1  
\end{array}\right].$$

In this example, we will not subdivide the files into stripes, and one sub-query is required in which we will decode both parts of the file. For this reason, we will remove the subscript for simplicity.
Similar to the scheme in section~\ref{sec:constructionb2}, the user generates $2$ random (column) vectors ${\mathbf{u}}_1$ and ${\mathbf{u}}_2$, of length $m$ each, whose elements are chosen uniformly at random from $GF(q)$. Define $U$ to be the $m \times 2$ matrix with its columns being the $b = 2$ random vectors $\mathbf{u}_1$ and $\mathbf{u}_2$, \emph{i.e.}, $$\underset{{m \times 2}}{U} = \left[ {\begin{array}{cc}
{{\mathbf{u}}_{1}} & {{\mathbf{u}}_{2}}
\end{array}} \right].$$

Now the user generates $m$ random codewords in the dual codes by multiplying the random matrix $U$ by the parity check matrix $\underset{b\times (b+k)}{H_1} $ to calculate \begin{equation}U {H_1} = \left[ \mathbf{q}'_1, \mathbf{q}'_2, \mathbf{u}_1, \mathbf{u}_2  \right],\end{equation}
and multiplying the random matrix $U$ by the parity check matrix $\underset{b\times (b+k)}{H_2} $ to calculate \begin{equation}
U {H_2} = \left[ \mathbf{q}'_3, \mathbf{q}'_4, \mathbf{u}_1, \mathbf{u}_2  \right],\end{equation}

The query vectors to nodes $1, 2, 3,$ and $4$ are chosen as follows:
\begin{equation}
	{\mathbf{q}}_{l} = 
	\begin{cases}
	{\mathbf{q}'_l}+\mathbf{e}_f, \quad\quad & \text{if } l = 1 \text{\quad mod $k$},\\
	{\mathbf{q}'_{l}}, \quad\quad & \text{otherwise.}
	\end{cases} 
\end{equation}

The query vectors to nodes $5,6$ are \begin{equation}
	{\mathbf{q}}_{l} = {\mathbf{u}_l}.
\end{equation}



\noindent\textit{Decodability:} 

Each of the punctured codes is coded as in section~\ref{sec:constructionb2}. Based on the decodability proved in section~\ref{sec:proof2}, the user can decode $a_f$ from the first code, and $a_f+b_f$ from the second code. Hence, the user retrieves file $X^f$.

\noindent\textit{Privacy:} 

The queries are sent by projecting the random matrix $U$ on the matrix

$$H = \left[ \begin{array}{ccc}
-P^T_1 & -P^T_2 & I
\end{array}\right].$$

This is a dual of the code generated by the matrix $$G = \left[ \begin{array}{ccc}
B_1 & 0 & P \\
0 & B_2 & P \\
\end{array}\right].$$

The dual code is an $(6,4,3)$ MDS code, and thus the queries are sent using a $(6,2,5)$ MDS code.

This means that any $2$ queries are linearly independent and thus if this is private against $2$ colluding nodes.

The user contacts $6$ nodes, to download $2$ information parts, thus the price of privacy of this is $cPoP = \frac{b+\delta k}{\delta} = \frac{6}{2}=3$.

\end{example}

\subsection{General Proof of Theorem~\ref{th:collusion2}}

We assume the user wants file $X^f$. Assume $n = b+\delta k$.{\footnote {If $b<n-\delta k$, only $b+\delta k$ nodes are queried.}} The user uses the last $b$ nodes, $n_{\delta k+1}, \dots, n_{\delta k+b}$ as common nodes. The rest of the nodes will be divided into $\delta$ groups, $j=1, \dots, \delta$, of $k$ nodes each.
This forms $\delta$ punctured $(b+k,k)$ MDS codes, each with a generator matrix $\left[ {\begin{array}{c|c}
	B_j & P_{b\times b} 
	\end{array}} \right]$ which can be transformed to systematic form $G_j = \left[ {\begin{array}{c|c}
	I_{k\times k} & P_j 
	\end{array}} \right]$ by multiplying by the inverse of the $k\times k$ matrix $B_j$.
Here we will use $\alpha = \delta$ subdivisions and $k$ queries.\footnote{ The  parameters can be optimized to  $\alpha=\frac{LCM(\delta,k)}{k}$ and $\rho= \frac{LCM(\delta,k)}{\delta}$, as was done in Example~\ref{ex:th2th32}. But to simplify notation, we will take $\alpha = \delta$ and $\rho=k$.}

We calculate the parity check matrix of the $\delta$ codes. 
$$ \underset{b\times (b+k)}{H_j} = \left[ {\begin{array}{c|c}
	-P_j^T & I_{b\times b}
	\end{array}} \right]. $$

Now for each sub-query, $i$, $i = 1,\dots, k$, the user generates $m$ random codewords by multiplying the random matrix $U_i = \left[\begin{array}{ccc}
\mathbf{u}_{1,i} & \dots & \mathbf{u}_{b,i}\\
\end{array}\right] \in GF(q)^{m \times b}$ by the parity check matrix $\underset{b\times (b+k)}{H_j}$ of subcode $j$ $$U_i \underset{b\times (b+k)}{H_j} = \left[ \mathbf{q}'_{1+(j-1)k,i}, \mathbf{q}'_{2+(j-1)k,i} \dots, \mathbf{q}'_{jk,i},\mathbf{u}_{1,i}, \dots, \mathbf{u}_{b,i} \right].$$

For the nodes $l=1, \dots, \delta k$, the query vectors in sub-query $i$ are as follows:
\begin{equation}
	{{\mathbf{q}}_{l,i}} = 
	\begin{cases}
	\mathbf{q}'_{l,i}+\mathbf{e}_{(f-1)\delta+j}, \quad\quad & \text{if } l = k-(j-1)+i,\\
	\mathbf{q}'_{l,i}, \quad\quad & \text{otherwise.}
	\end{cases} 
\end{equation}

For the nodes $l=\delta k+1, \dots, \delta k+b$, the query vectors in sub-query $i$ are the columns of $U_i$

\begin{equation}
	\mathbf{q}_{l,i} = \mathbf{u}_{l-\delta k,i}.
\end{equation}


\noindent\textit{Decodability:} 

For each subcode $j$, we follow the scheme of Theorem~\ref{th:collusion} to obtain the $j^{th}$ stripe of file $x^f$. Subsequently, the user is able to decode the file $x^f$.





\noindent{\em Privacy:} The queries are generating by multiplying the random matrix $U$ on the matrix

$$H = \left[ \begin{array}{ccccc}
-P^T_1 & -P^T_2 & \cdots & -P^T_{\delta} & I
\end{array}\right].$$

This is a dual of the code generated by the matrix $$G = \left[ \begin{array}{ccccc}
B_1 & 0 & \cdots & 0 & P \\
0 & B_2 & \cdots & 0 & P \\
\vdots & \vdots & \ddots & \vdots & \vdots \\
0 & 0 & \cdots & B_\delta & P \\
\end{array}\right].$$

We see that the code generated by $G$ is an $(\delta k+b,\delta k, b+1)$ MDS code, and thus the queries are sent using an $(\delta k+b,b,\delta k+1)$ MDS code.

This means that any $b$ queries are linearly independent and thus this is private against $b$ colluding nodes.

The user contacts $b+\delta k$ nodes, to download $\delta$ information parts, thus the price of privacy of this is $cPoP = \frac{b+\delta k}{\delta}$.

\section{Comparison to fundamental bounds}

Our scheme achieves the fundamental bounds currently known for infinite number of files and $1$ spy node, i.e. no collusion.
The lowest achievable price of privacy of a storage system with replicated databases is given in \cite{sun2016capacity} to be $\frac{1-(1/n)^m}{1-(1/n)}$ which asymptotically approaches $\frac{n}{n-1}$ as $m\rightarrow \infty$. If we apply our PIR scheme for a replicated database, the $cPoP=\frac{1}{1-R} = \frac{n}{n-1}$ which is the limit of  the lower bound.

The lower bound for an $(n,k,d)$ MDS-coded database was derived in \cite{banawan2016capacity} to be $\frac{1-(k/n)^m}{1-(k/n)}$, which asymptotically approaches $\frac{n}{n-k}=\frac{1}{1-R}$ as $m\rightarrow\infty$, and is again the cPoP achieved by our construction. 

\section{Conclusion}\label{sec:conc}

We studied the problem of constructing  PIR schemes with low  communication cost for requesting data from a DSS that uses  MDS codes. Some nodes in the  DSS may be spies   who will report to a third party, such as an oppressive regime, which  data is being requested by a  user. The objective  is to allow the user to obtain its requested data without revealing any information on the identity of the data to the nodes. We constructed PIR schemes against non-colluding nodes that achieve the information theoretic limit on the download communication cost for linear schemes. An important property of these schemes is their universality since  they depend on the code rate, but not on the MDS code itself. 
When there is $b$-collusion with $2 \leq b \leq n - k$, we devised linear PIR schemes that have download cost equal to $b + k$ per unit of requested data.


\section{Appendix}\label{sec:app}

\subsection{Alternative Proof of Theorem~\ref{th:collusion}}\label{sec:proof2A}

To simplify the description of the scheme,  we will assume $b=n-k$.
The scheme has dimension $\rho=k$, \emph{i.e.}, it consists of $\rho=k$ sub-queries. Moreover, the scheme requires no subdivisions, \emph{i.e.}, the number of stripes  $\alpha = 1$. Since there are no subdivisions, we simplify further the notation and write $x_{i1}^j = x_i^j$ to denote the $i^{th}$ systematic symbol of file $X^j$, where $j = 1, \dots, m$.
Denote by $f$ the index of the file that  the user wants, \emph{i.e.}, the user wants to retrieve file $X^f$.

In the $i^{th}$ sub-query, $i=1,\dots,k$, the proposed PIR scheme retrieves systematic symbol  $x_i^f$ of the wanted file $X^f$.
 So, by the completion of the scheme, the user will have obtained all the $k$  symbols forming the file.  
 
In sub-query $i$, the user creates $b$ random (column) vectors ${{\mathbf{u}}_{1,i}}, \dots, {{\mathbf{u}}_{b,i}}$, of dimension $m$ each, whose elements are chosen uniformly at random from $GF(q)$. Define $U_i$ to be the $m \times d-1$ matrix with its rows being the $b = n-k$ random vectors used in sub-query $i$, \emph{i.e.}, $$\underset{{m \times d-1}}{U_i} = \left[
   \mathbf{u}_{1,i}, 
   \mathbf{u}_{2,i}, 
   \dots, 
   \mathbf{u}_{d-1,i} \right].$$
   
Recall that the generator matrix of the MDS code is $$\underset{k\times n}{\Lambda} = \left[ {\begin{array}{c|ccc}
\multirow{3}{*}{$I_{k\times k}$} & \lambda_{1,k+1} & \dots & \lambda_{1,n} \\
 & \vdots & \vdots & \vdots \\
 & \lambda_{k,k+1} & \dots & \lambda_{k,n} 
\end{array}} \right].$$
We write $\Lambda = \left[ {\begin{array}{c|c}
I & P
\end{array}}\right]$, where 

$$\underset{k\times d-1}{P} =  \left[ {\begin{array}{ccc}
\lambda_{1,k+1} & \dots & \lambda_{1,n} \\
  \vdots & \vdots & \vdots \\
  \lambda_{k,k+1} & \dots & \lambda_{k,n} 
\end{array}} \right].$$

We denote by ${\mathbf{p}}^T_j$ the $j^{th}$ row of $P$.  
Let ${\mathbf{e}}^T_f = \left[ {\begin{array}{ccc}
\mathbf{0_{1\times (f-1)}} & 1 & \mathbf{0_{1\times (m-f)}}
\end{array}} \right]$.

 
For a systematic node $l$, the user sends the sub-query vector:

\begin{equation}
{\mathbf{q}}_{l,i} =
\begin{cases}
\lambda_{l,k+1}{\mathbf{u}}_{1,i} + \dots + \lambda_{l,n}{\mathbf{u}}_{d-1,i} + {\mathbf{e}}_f, & \text{if } l = i, \\
\lambda_{l,k+1}{\mathbf{u}}_{1,i} + \dots + \lambda_{l,n}{\mathbf{u}}_{d-1,i}, & \text{otherwise.}
\end{cases}
\end{equation} 

This translates to

\begin{equation} 
{{\mathbf{q}}_{l,i}} = 
\begin{cases}
{U_i}{\mathbf{p}}_l + {\mathbf{e}}_f, \quad\quad & \text{if } l = i,\\ \label{eq:q1}
{U_i}{\mathbf{p}}_l, \quad\quad & \text{otherwise.}
\end{cases} 
\end{equation} 


For the parity nodes $l=k+1, \dots, n=k+b$, the $i^{th}$ sub-query vector is given by,
\begin{equation}
{{\mathbf{q}}_{l,i}} = {\mathbf{u}}_{l-k, i}. \label{eq:q2}
\end{equation}

Therefore, the  response of node $l$ to the $i^{th}$ sub-query, denoted by ${r}_{l,i}$,  is given by \eqref{eq:r1} and can be written as
\begin{equation}\label{eq:res}
{r}_{l,i} = {{\mathbf{q}}_{l,i}}^T {\mathbf{w}}_l,
\end{equation}
where ${\mathbf{w}}_l$ is the vector representing the data stored on node $l$.



We prove Theorem~\ref{th:collusion} by showing that the scheme described in Section~\ref{sec:constructionb2} ensures decodability and privacy. The main ingredient in the proof, which makes it different from   the proof of Theorem~\ref{th:main},  is that the scheme does not require the user to decode all the interference terms.

%
%


Recall that the user wants to retrieve file $X^f$. We will prove that the user can retrieve $x^f_i$ in the $i^{th}$ sub-query. 


\textit{Decodability:} 

From  \eqref{eq:q1} and \eqref{eq:res}, the response of systematic node $l$ to the $i^{th}$ sub-query is given by
\begin{equation}
{r}_{l,i} = 
\begin{cases}
{\mathbf{p}}_l^T U_i^T {\mathbf{w}}_l+x_i^f = {\mathbf{w}}_l^T {U_i} {\mathbf{p}}_l + x^f_i & \text{if } l=i, \\
{\mathbf{p}}_l^T U_i^T {\mathbf{w}}_l = {\mathbf{w}}_l^T {U_i} {\mathbf{p}}_l & \text{otherwise.}
\end{cases}
\end{equation}

Notice that ${\mathbf{w}}_l^T{U_i}^T{\mathbf{p}}_l$ is the $l^{th}$ diagonal element of $\mathscr{X} {U_i}^TP^T$, since ${\mathbf{w}}_l$ is the $l^{th}$ row of  $\mathscr{X}, l=1,\dots,k,$ due to the  assumption that the MDS code is systematic.
Thus, the vector representing all  the responses of the systematic nodes to the $i^{th}$ sub-query can be written as follows,
\begin{equation}
\left[ {\begin{array}{c}
{r}_{1,i}\\
{r}_{2,i}\\
\vdots \\ 
r_{k,i} \end{array}}\right] = \text{diag}(\mathscr{X}^T {U_i}P^T) + \left[ {\begin{array}{c}
\mathbf{0_{i-1\times 1}} \\
x^f_i\\
\mathbf{0_{k-i\times 1}} \end{array}}\right]  \label{eq:s1},
\end{equation}
where diag($\cdot$) is the diagonal of the corresponding matrix.

Denoting by ${\mathbf{p}}'_{j}$ the $j^{th}$  column of $P$, the response of parity node $l,l=k+1,\dots,n,$ can be written as
\begin{align}
r_{l,i} & = {\mathbf{u}}^T_{l-k,i}{\mathbf{w}}_l\\
& = {\mathbf{w}}_l^T {\mathbf{u}}_{l-k,i} \\
&= {{\mathbf{p}}'}^T_{l-k} \mathscr{X}^T {{\mathbf{u}}_{l-k,i}},\label{eq:res}
\end{align}
where \eqref{eq:res} follows from the fact that the coded data stored on parity node $l$ can be written as ${\mathbf{w}}_l={{\mathbf{p}}'}^T_{l-k} \mathscr{X}$.
Thus, similarly to~\eqref{eq:s1}, we  can write all the responses of the parity nodes in vector form as 
\begin{equation}\label{eq:p}
\left[ {\begin{array}{c}
r_{k+1,i}\\
r_{k+2,i}\\
\vdots \\ 
r_{n,i} \end{array}}\right] = \text{diag}(P^T \mathscr{X}^T{U_i}^T).
\end{equation}

Next, we want to show that $x_i^f$ can be decoded as follows,
\begin{equation*}
x_i^f = \sum_{l=1}^k r_{l,i} - \sum_{l=k+1}^{k+b} r_{l,i}.
\end{equation*}

Indeed, we have 
\begin{align}
\sum_{l=1}^k r_{l,i} &= \tr(\mathscr{X}^T {U_i}P^T) + x_1^f \label{eq:d0}\\ \label{eq:d1}
&= \tr(P^T \mathscr{X}^T{U_i}) + x_1^f \\
&=\sum_{l=1}^{b} r_{l+k,i} + x_1^f \label{eq:d2}\\
&=\sum_{l=k+1}^{k+b} r_{l,i} + x_1^f,
\end{align}
where $\tr(\cdot)$ is the trace operator, \eqref{eq:d0} follows from \eqref{eq:s1},  \eqref{eq:d1} follows from the trace  property, $\tr(A B C) = \tr(C A B)$, and \eqref{eq:d2} follows from \eqref{eq:p}. 


This means that the responses of the systematic nodes and those of the parity nodes   cancel out to leave the part required, \emph{i.e.}, $x_i^f$. Therefore, we showed that in the $i^{th}, i=1,\dots,k$ sub-query, the user can decode $x_i^f$ and by the completion of the $k^{th}$ sub-query the user would have obtained   the whole file $X^f$.

\bibliographystyle{ieeetr}
\bibliography{coding2,coding1,}

%
%
%
%
%
%
\end{document}